\documentclass[aps,pra,twocolumn,superscriptaddress,showpacs,floatfix]{revtex4-2}
\long\def\/*#1*/{}
\usepackage{amsfonts}
\pdfoutput=1
\usepackage{graphicx}
\usepackage{amsmath}
\usepackage{amsthm}
\usepackage{comment}
\usepackage[colorlinks=true, citecolor=blue, urlcolor=blue ]{hyperref}
\usepackage{graphicx}
\usepackage{braket}
\usepackage[author={AB}]{pdfcomment}
\usepackage{graphicx}
\usepackage{subcaption}
\usepackage{float}    % enables [H]
\usepackage{caption}
\usepackage{ragged2e}

\begin{document}
\title{Deterministic distribution of W-class states in quantum networks}
\author{Souvik Chatterjee}
\email{souvik19n@gmail.com}
\affiliation{Center for Quantum Engineering, Research, and Education, TCG CREST, Bidhan Nagar, Kolkata - 700091, India.}
\author{Prasenjit Deb}
\email{devprasen@gmail.com}
\affiliation{Center for Quantum Engineering, Research, and Education, TCG CREST, Bidhan Nagar, Kolkata - 700091, India.}
\author{Chandan Datta}
\email{dattachandan10@gmail.com}
\affiliation{Department of Physics, Indian Institute of Technology Jodhpur, Jodhpur 342030, India}
\author{Pankaj Agrawal}
\email{pankaj.agrawal@tcgcrest.org}
\affiliation{Center for Quantum Engineering, Research, and Education, TCG CREST, Bidhan Nagar, Kolkata - 700091, India.}

\date{\today}
%%%%%%%%%%%%%%%%%%%%%%%%%%%%%%%%%%%%%%%%%%%%%%%%

\begin{abstract}
Multipartite entangled states possess a number of non-intuitive properties, making them a useful resource for various quantum information-processing tasks. The three-qubit W-state is one such example where every state is robust to single-qubit loss. However, this state is not suitable for deterministic distribution, and deterministic communication protocols.  Here, we focus on the distribution of a non-symmetric version of such states, namely $W_{\mathrm{mod}}$ states. These states belong to the W-class, and have one ebit of entanglement across a specific bipartition, enabling deterministic teleportation and superdense coding. In particular, we describe a few protocols through which these multipartite entangled states can be distributed {\it deterministically} in a quantum network by first preparing them locally in a central node and then transmitting individual qubits to the end nodes. We analyse the performance of these protocols based on the fidelity of the final distributed state, considering all types of noises that can act during the distribution. Finally, we compare the performance of the protocols to the case where the distribution is performed without any central node.   
\end{abstract}
\maketitle

\section{Introduction}\label{sec1}

Quantum networks\cite{kimble2008,Sasaki_2017,Tajima_2017} are likely to play an important role in everyday life in near future.
%in are instrumental in the development of quantum technologies. 
In recent years, they have attracted significant research interest due to their wide range of promising applications, including secure communication\cite{Tajima_2017}, exponential advantages in communication complexity\cite{communication_complexity}, distributed quantum computation\cite{dist_comp}, and distributed quantum sensing \cite{qnet_sensing}. So far, most studies have focused on distributing bipartite entangled states, particularly Bell states, as they offer optimal performance in key quantum information processing tasks such as quantum key distribution \cite{E91,qkd_bb84,qkd}, quantum teleportation \cite{teleportation, tele_experiment}, and blind quantum computation. However, the practical realization of full-fledged quantum network architectures and protocols requires the reliable generation and distribution of multipartite entangled states across quantum networks. 
\paragraph*{}
Recent developments of quantum networks have led to several important milestones including the distribution of entanglement over 1200 kilometers using a satellite \cite{1200}, quantum teleportation without using a pre-shared entangled state \cite{teleportation_without_ent}, the generation of light-matter entanglement over 50 kilometers of optical fiber through the use of quantum frequency conversion \cite{LM_entanglement}, and the creation of the first three-node quantum network \cite{3node_network}. Given the condition that distributing multipartite entangled states is critical to exploit the new technology of quantum networks, one fundamental question naturally arises: how efficiently multipartite entangled states can be distributed over quantum networks? Standard approaches often rely on preparing the target state locally in a node, generally termed as \textit{central node}, and then transmitting each qubit of the multipartite state to individual parties (end nodes). The distribution of the entangled qubits from central node to end nodes is not a unique process. The distribution can take place in many different ways, such as direct transmission through quantum channels, teleportation of the entangled qubits using preshared Bell pairs between the end nodes and the central node, and so on. For example, in the case of a symmetric star-shaped network, depending on the availability of resources, the central node can deploy unique independent strategies \cite{s_wehner} to distribute the multipartite entangled state across to the end nodes.  Recent research have demonstrated distribution of $N$-partite Greenberger-Horne-Zeilinger (GHZ)\cite{s_wehner} and other graph states\cite{graph_state_dist} in various types of quantum networks using the central-node architecture. Moreover, robustness of distributed entangled states against different types of noises, such as noisy connections, noisy memory, and noisy Bell-state measurements (BSM), has also been studied in \cite{s_wehner}. Though the distribution of GHZ states in quantum networks has received much attention, both theoretically and experimentally, The distribution of W states is underexplored. 
\paragraph*{}
In this article, we focus on deterministic distribution of W-class states. Due to their ability to retain bipartite entanglement after loss of a single qubit, W states may be effective alternative to GHZ states for quantum information processing tasks \cite{Joo,PhysRevA.98.052320,ShiTomita2002,D'Hondt}. However, the canonically symmetric W states are not suitable for unit-fidelity teleportation with unit probability and deterministic superdense coding. Nevertheless, there exists a class of \emph{non-symmetric} W states that belongs to W-class \cite{dur_W} and are suitable for deterministic communication \cite{modified_W_pati}. These states possess genuine tripartite entanglement and have one ebit of entanglement across a specific bipartition, enabling deterministic quantum information processing tasks \cite{Agrawal:2015wfz} while maintaining loss tolerance like W states \cite{modified_W_pati}. The states are

\begin{equation} \label{eq0}
|W_m\rangle_{123} = \frac{1}{\sqrt{2 + 2m}} (|100\rangle + \sqrt{m}e^{i\gamma}|010\rangle + \sqrt{m + 1}e^{i\delta}|001\rangle),
\end{equation}

where $m$ is a real number and $\gamma~\text{and}~ \delta$ are phases. These phases can be removed by local unitary transformations. We consider the states for which $m = 1$ and the phases are set to zero , {\it i.e.},
\begin{equation}\label{eq1}
\ket{W_{\mathrm{mod}}}_{123} = \frac{1}{2}\ket{100} + \frac{1}{2}\ket{010} + \frac{1}{\sqrt{2}}\ket{001}.
\end{equation}
In our work, we consider the central-node architecture of quantum networks and describe different distribution protocols for these states. Later, we consider various types of noise and analyse the performance of each protocol. Finally, we qualitatively compare the performance of the central-node based protocols with the other protocols for distributing W states.

The rest of the article is arranged as follows: in Section (\ref{sec2}), we describe the properties of $W_{\mathrm{mod}}$ states. Section (\ref{sec3}) describes a few distribution protocols for the states in a quantum network. In Section (\ref{sec4}), we analyse the protocols in noisy scenarios. We compare the protocols mentioned in this article with the other protocols for distributing W states in Section (\ref{sec5}). Finally, in Section (\ref{sec6}), we present conclusions. 

%%%%%%%%%%%%%%%%%%%%%%%%%%%%%%%%%%%%%%%%%%%%%%%%%%%%%%%%%
\section{Properties of $W_{\mathrm{mod}}$ states and their application in quantum information processing tasks}\label{sec2}

It is known that under stochastic local operations and classical communication (SLOCC), three-qubit pure multipartite entangled states can be divided into two classes -- GHZ and W \cite{dur_classification}. The GHZ class of states exhibit nonzero three-tangle and lose all bipartite entanglement when any of the qubits is traced out. Whereas, W class of states have zero three-tangle and nonzero pairwise concurrence \cite{three_tangle}. This retention of bipartite entanglement under single qubit loss makes these states an important resource for networks having lossy channels or unreliable end nodes. The canonical W state is
\begin{equation}
\ket{W}=\frac{1}{\sqrt{3}}\left(\ket{100}+\ket{010}+\ket{001}\right).
\end{equation}
This state is robust against qubit loss but does not qualify as a resource for deterministic teleportation or standard superdense coding \cite{modified_W_pati}. However there exists a family of W states \cite{modified_W_pati} with tailored amplitudes that exhibits one ebit of entanglement across a specific bipartition and can be used for unit-fidelity teleportation with unit probability and superdense coding. Such states are termed as \textit{modified} W states, $W_{\mathrm{mod}}$, as given in Eq (\ref{eq1}). From quantum network perspective, the third qubit of the state can play the role of a \emph{routing} qubit, whereas, qubits~1 and~2 can be client qubits.

%%%%%%%%%%%%%%%%%%
\subsection{Preparation of $W_{\mathrm{mod}}$ states and their entanglement properties}

A $W_{\mathrm{mod}}$ state, given in Eq. \ref{eq1}, can be deterministically generated by applying a two-qubit unitary between a Bell state and an ancilla qubit. Specifically, starting with a product state of the form
\begin{equation}\label{eq3}
|\phi^+\rangle_{13}\otimes|0\rangle_2 = \frac{1}{\sqrt{2}}\big(|000\rangle + |101\rangle\big),
\end{equation}
the action of the unitary
\begin{equation}
U_{MW} =
\begin{bmatrix}
0 & 0 & 1 & 0 \\
\tfrac{1}{\sqrt{2}} & 0 & 0 & \tfrac{1}{\sqrt{2}} \\
\tfrac{1}{\sqrt{2}} & 0 & 0 & -\tfrac{1}{\sqrt{2}} \\
0 & 1 & 0 & 0
\end{bmatrix}
\end{equation}
on qubits $(1,2)$ yields,
\begin{equation}
U_{MW}(|\phi^+\rangle_{13}\otimes|0\rangle_2)
= \tfrac{1}{2}|100\rangle + \tfrac{1}{2}|010\rangle + \tfrac{1}{\sqrt{2}}|001\rangle,\nonumber\\
\end{equation}
which is exactly the desired $W_{\mathrm{mod}}$ state. In a similar way, a complete $W_{\mathrm{mod}}$-basis set can be constructed starting from the state mentioned in Eq (\ref{eq3}). Consider the following unitary acting on qubits (2,3) of the state,
\begin{equation}
U_{MWB} =
\begin{bmatrix}
0 & 1 & 0 & 0 \\[4pt]
\tfrac{1}{\sqrt{2}} & 0 & 0 & \tfrac{1}{\sqrt{2}} \\[4pt]
\tfrac{1}{\sqrt{2}} & 0 & 0 & -\tfrac{1}{\sqrt{2}} \\[4pt]
0 & 0 & 1 & 0
\end{bmatrix}. \nonumber\\
\end{equation}
The resulting state will be,
\begin{eqnarray}
U_{MWB}(|\phi^+\rangle_{13}\otimes|0\rangle_2) 
&=& \tfrac{1}{2}|010\rangle + \tfrac{1}{2}|001\rangle + \tfrac{1}{\sqrt{2}}|100\rangle \nonumber\\
&=& |\eta^{+}\rangle,
\end{eqnarray}
which is one of the basis states. Now applying local unitary on qubit~1 of this state, the remaining three states $|\eta^-\rangle,|\xi^+\rangle,~ \text{and}~ |\xi^-\rangle$ of the set can be obtained as -- 
\[
|\eta^-\rangle = Z_1|\eta^+\rangle,\qquad
|\xi^+\rangle = X_1|\eta^+\rangle,\qquad
|\xi^-\rangle = X_1Z_1|\eta^+\rangle.
\]
where, $X$ and $Z$ are the Pauli operators.
Thus, the  orthonormal set $\{|\eta^\pm\rangle,|\xi^\pm\rangle\}$ can be realized from a single preparation circuit followed by simple, classically controlled single-qubit unitary operations. This construction provides a practical method for implementing $W_{\mathrm{mod}}$-basis measurements in network protocols. If we analyse the entanglement properties of $W_{\mathrm{mod}}$ states, we find the following:

\begin{itemize}
    \item For any state $\ket{W_{\mathrm{mod}}}$,  three-tangle $\tau_{123}=0$.
    
    \item The reduced state of qubit~3 is $\rho_3=\tfrac{1}{2}\mathbb{I}$ and the corresponding von Neumann entropy $S(\rho_3)=1$. Therefore, $W_{\mathrm{mod}}$ states have one ebit of entanglement for the bipartition $\rho_{12|3}$, where $\rho$ is the density matrix corresponding to the state $\ket{W_{\mathrm{mod}}}$.
\end{itemize}
The second property mentioned above clearly explains why quantum information processing tasks that requires one ebit of bipartite entanglement, such as teleportation and dense coding, can be implemented deterministically using these states.

%%%%%%%%%%%%%%%%%%%%%%%%%
\subsection{Deterministic teleportation and superdense coding using $W_{\mathrm{mod}}$ states}
Quantum information processing tasks, such as deterministic teleportation and superdense coding, require one ebit of entanglement shared between two parties. Following the construction in \cite{modified_W_pati}, the sender and receiver first share a $W_{\mathrm{mod}}$ state between them in such a way that the qubits~1 and~2 are in possession of the sender and qubit~3 is at the receiver's lab, ensuring that they share one ebit of entanglement between them. To teleport an unknown quantum state, the sender performs a suitable three-qubit projective measurement in a $W_{\mathrm{mod}}$-basis and communicate the two classical bits to the receiver, who applies a unitary on his qubit to recover the original unknown qubit state with unit fidelity. For superdense coding, the sender first applies local unitaries on qubit~1, producing four orthogonal W-class states. Later, he sends one of his qubit to the receiver and the receiver performs appropriate three-qubit measurements to extract the information that the sender has encoded.

Having described the properties and usefulness of $W_{\mathrm{mod}}$ states, we now describe various distribution protocols for such states in the next section.

%%%%%%%%%%%%%%%%%%%%%%%%%%%%%%%%%%%%%%%%%%%%%%%%%%%%%%%%%%%%%%%%%%%%%%%%%%
\section{Distribution protocols}\label{sec3}
The distribution of W states in different network scenarios have been considered \cite{Qrepeaters_W}.We consider an asymmetric star-shaped network with a central-node and describe possible strategies for distributing 3-qubit $W_{\mathrm{mod}}$ states. As these states exhibit one ebit of entanglement only across a certain bipartition, their distribution using the central-node architecture will significantly differ from that of GHZ states. 

\subsection*{Protocol 1: Direct transmission of entangled qubits}\label{protocol1}

%%%%%%%%%%%%%%%%%%%%%%%%%%%%
\begin{figure}[h]
    \centering
    \includegraphics[width= \linewidth]{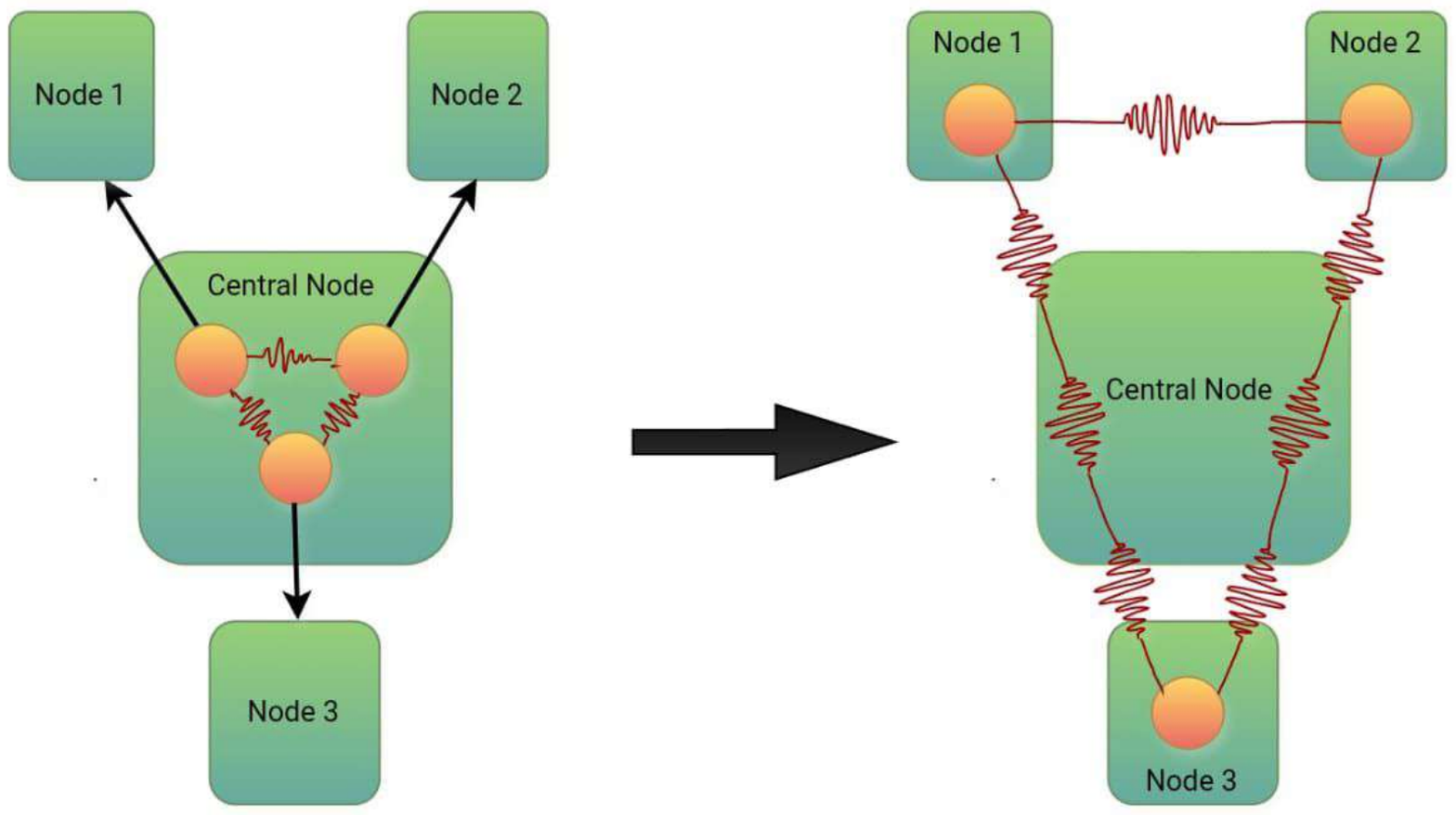}
    \caption{\justifying{Schematic diagram showing direct distribution of a $W_{\mathrm{mod}}$ state. The central node has no memory. It locally prepares the target state and then sends each qubit through direct transmission to the end nodes.}}
    \label{fig:1}
\end{figure}

%%%%%%%%%%%%%%%%%%%%%%%%%%%%

In a star-shaped network, the simplest method to distribute any multipartite entangled state is to generate it locally in the central node and then transmitting the entangled qubits to the end nodes (FIG.~\ref{fig:1}). This method is suitable for such a network where the central node does not have any quantum memory. This memoryless protocol requires connections between the end nodes and central node through which qubits (the information carriers) can be directly transmitted. (For photonic qubits, optical fibers may constitute the transmission lines.)  Apart from this method, there is another scheme using which a multipartite entangled state can be distributed in a quantum network having memoryless central node. In this scheme, the end nodes emit entangled qubits that are sent through the connections to the central node. The qubits are then interfered and measured in the $W_{\mathrm{mod}}$ basis followed by classical communication, resulting in the creation of the target state on the end nodes. These type of schemes have been demonstrated for the distribution of GHZ states \cite{s_wehner} and W states \cite{Qrepeaters_W}. Compared to other distribution schemes, direct ones share the advantage of having a central node that can be very simple. In case of photonic qubits, one may require only linear-optics components and single-photon detectors. However, a major disadvantage of these schemes is that their success relies on the simultaneous arrival of all photons at the central node, making them very sensitive to photon losses. The scaling of this type of direct entanglement distribution can be determined from the transmittance $\eta$ of the connections. If there are $N$ number of end nodes in the network and each node transmits photons successfully with probability $\eta$, then the distribution rate will scale as $\eta^{N}$ \cite{s_wehner}. 
%%%%%%%%%%%%%%%%%%%%%%%%%%%%%%%%%%%%%%
\subsection*{Protocol 2: Distribution through teleportation of entangled qubits}\label{protocol2}

%%%%%%%%%%%%%%%%%%%%%%%%%%%%
\begin{figure}[h]
    %\centering
    \includegraphics[width= 1.05
    \linewidth]{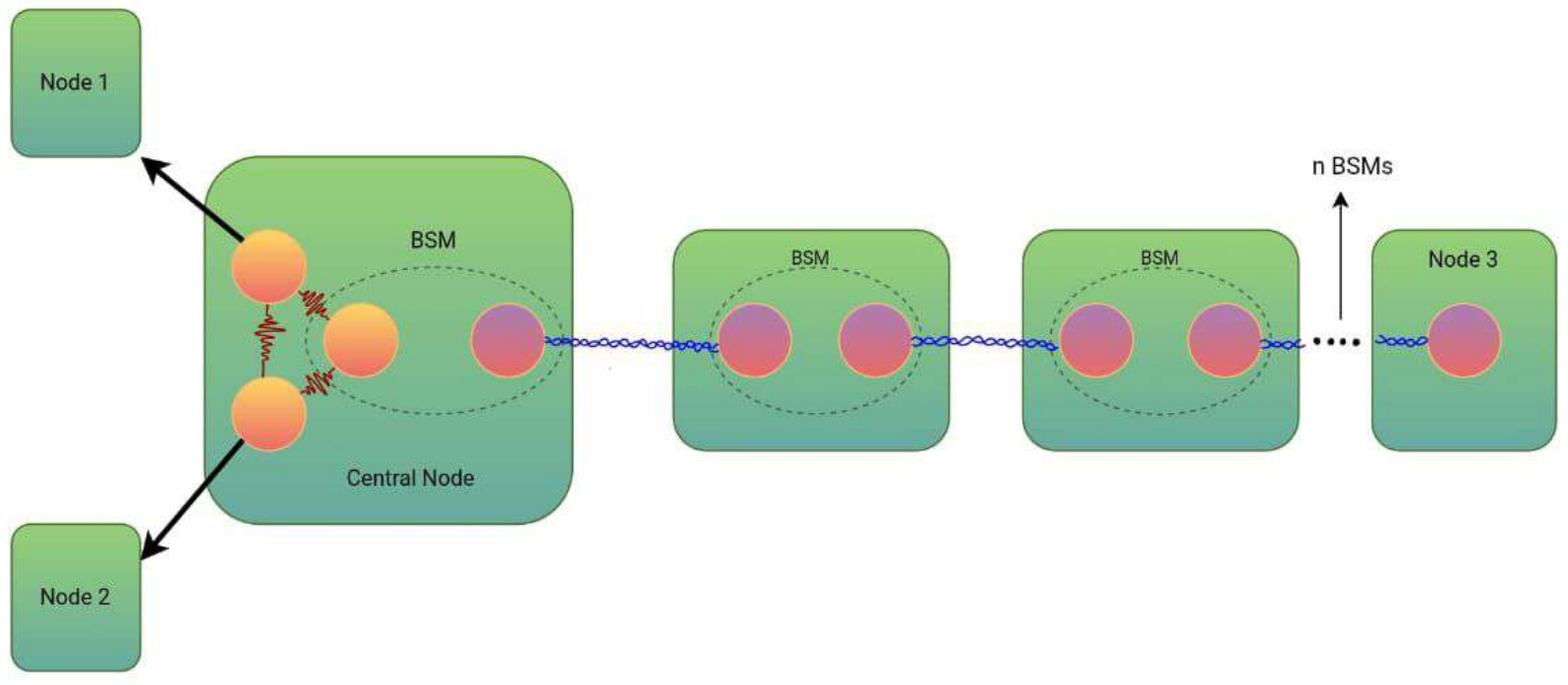}
    \caption{\justifying{Distribution of a $W_{\mathrm{mod}}$ state using a central node and a chain of repeaters. The central node first locally prepares the target state. Then Bell states are shared starting from the central node to the end node, where qubit 3 will be sent. The central first directly sends two qubits to two end nodes. Later through consecutive Bell measurements at the central and each repeater nodes, qubit 3 is sent to the end node, and the final target state is shared between the end nodes.}}
    \label{fig:2}
\end{figure}

%%%%%%%%%%%%%%%%%%%%%%%%%%%%
Apart from direct transmission of entangled qubits, a central node can also be differently used for distributing entanglement in a quantum network. More specifically, the multipartite entangled state can first be prepared locally in the central node and then the qubits can be transmitted to the end nodes through quantum teleportation using pre-shared Bell states \cite{Nielsen:2012yss}.The teleportation of the qubits is realized by performing a Bell-state measurement on the to-be-teleported qubit and a qubit in the Bell state. In \cite{s_wehner}, the authors have analysed the performance, namely, rate and fidelity, of such a protocol for a scenario where an $N$-qubit GHZ state is distributed in a symmetric star-shaped network. Here, we consider distribution of a 3-qubit $W_{\mathrm{mod}}$ state using this method. From the construction of $W_{\mathrm{mod}}$ states, it is clear that these states lack the symmetry present in GHZ states. Hence, the distribution of these states using entanglement swapping will differ significantly from that of GHZ states.  
The protocol is as follows (FIG.~\ref{fig:2}):
\begin{enumerate}
    \item Prepare a $\ket{W_{\mathrm{mod}}}$ in the central node. 
    \item Distribute a Bell state between the central node and one of the end nodes through repeated attempt.
    \item Perform a BSM at the central node between the qubit that holds part of the W state and qubit that holds part of the shared Bell state. 
    \item Communicate the measurement outcome to the end node over a classical channel.
    \item The end node performs single-qubit unitary operation on the qubit that it holds depending on the outcome of the BSM.
    \item Send the remaining two qubits to the other two end nodes.
\end{enumerate}

After the Pauli operation at one of the end nodes and direct transmission of the other qubits to the corresponding two nodes, the $W_{\mathrm{mod}}$ state is shared between the three end nodes. By coupling the memory qubit of the $W_{\mathrm{mod}}$ state to auxiliary Bell pairs and performing successive BSMs, the multipartite entanglement can be deterministically swapped across multiple nodes while preserving the functional form of the state. After $k$ hops, the qubits at the two end nodes and the final memory qubit at the distant node share the state
\begin{equation}
|\psi_{12R}\rangle = \tfrac{1}{2}|100\rangle_{12R} + \tfrac{1}{2}|010\rangle_{12R} + \tfrac{1}{\sqrt{2}}|001\rangle_{12R},
\end{equation}
demonstrating that the original structure is perfectly maintained with the memory qubit effectively shifting to the remote node. Here, qubits 1 and 2 are the transmitted qubits and qubit $R$ is the memory qubit. Unlike GHZ states, which are fragile under qubit loss, the $W_{\mathrm{mod}}$ state structure ensures that the bipartite entanglement between the two transmitted qubits persists and that entanglement swapping does not destroy the two-qubit entanglement. Furthermore, due to the symmetry of Bell states, all single-qubit depolarizing noise introduced during the intermediate swaps can be equivalently ``pushed'' to the end nodes, significantly simplifying the noise analysis. This property highlights $W_{\mathrm{mod}}$ states as a natural resource for scalable and fault-tolerant entanglement distribution in multi-hop quantum networks. An explicit derivation of this swapping-based distribution process of $W_{\mathrm{mod}}$ states is provided in the Appendix \ref{appendix A1}.
%%%%%%%%%%%%%%%%%%%%%%%%%%%%%%%%%%%
\subsection*{Protocol 3: Distribution through multipartite joint-measurement}\label{protocol3}

%%%%%%%%%%%%%%%%%%%%%%%%%%%%
\begin{figure}[t]
    \centering
    \includegraphics[width= \linewidth]{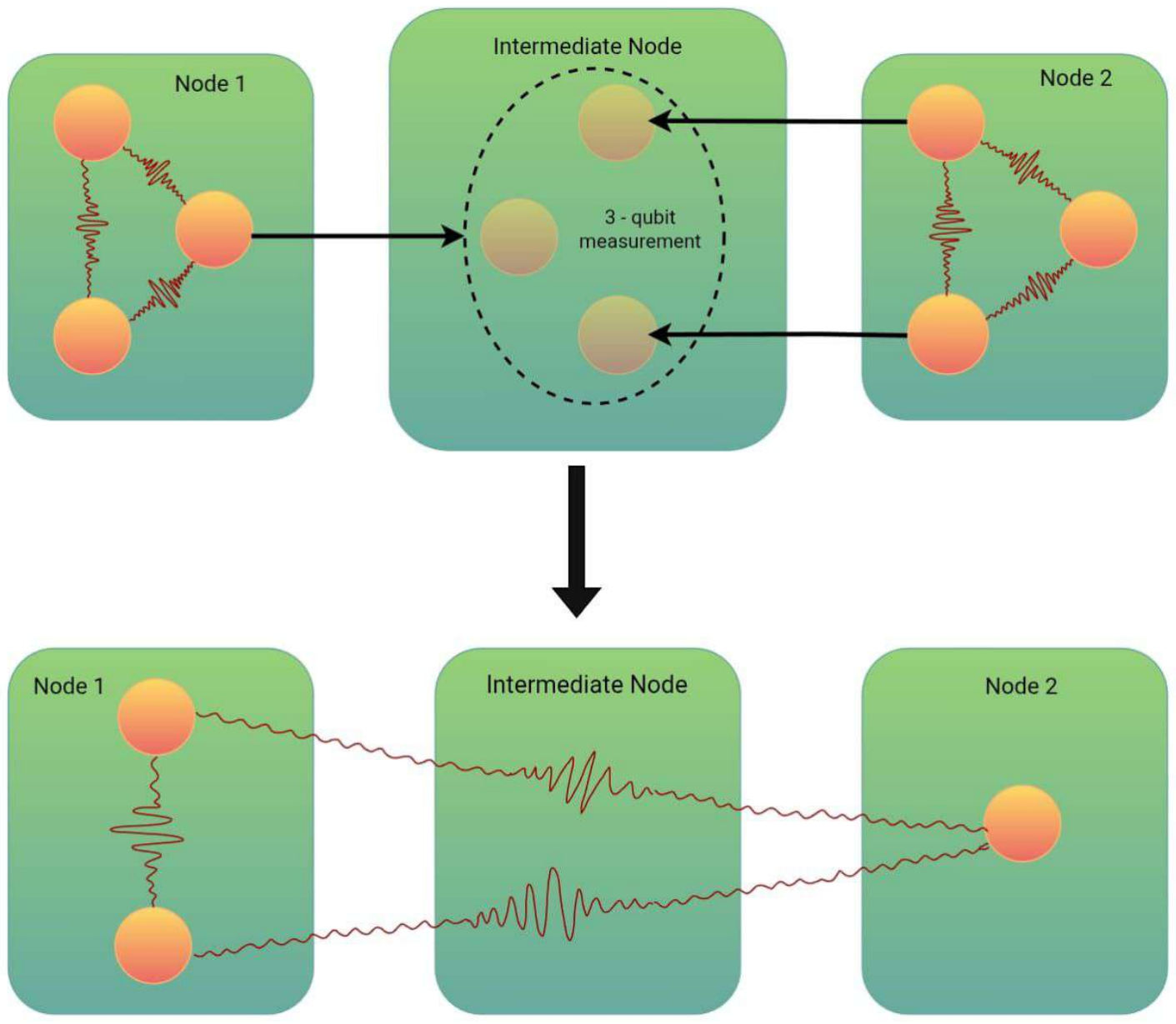}
    \caption{\justifying{A $W_{\mathrm{mod}}$ state can be distributed between two central nodes using an intermediate node in between them. The two central nodes first locally prepare two $W_{\mathrm{mod}}$ states. Then one of the central nodes sends two qubits and another one sends one qubit of their respective $W_{\text{mod}}$ states to the intermediate node. A joint three-qubit measurement is performed at the intermediate node and the target state is distributed between the central nodes.}}
    \label{fig:3}
\end{figure}

%%%%%%%%%%%%%%%%%%%%%%%%

An alternative distribution protocol to BSM-based swapping is a multipartite joint-measurement on parts of two $W_{\mathrm{mod}}$ states prepared at two different central nodes. The preparation of the states are done in such a way that for the state $|\psi\rangle_{123}$ at one central node $C$ state, the von Neumann entropy of the subsystems for the bipartition $\rho_{12|3}$ is unity, i.e., $S(\rho_{12}) = S(\rho_{3}) = 1$. Similarly, for the another $W_{\mathrm{mod}}$ state $|\psi\rangle_{456}$ prepared at a different central node $C^{\prime}$, the von Neumann entropy of the subsystems are $S(\rho_{45
}) = S(\rho_{6}) = 1$ for the bipartition $\rho_{45|6}$. In such a scenario, a $W_{\mathrm{mod}}$ state can be distributed across three end nodes by sending directly the qubits~1, 2, and 6 to those nodes and performing a joint measurement on the qubits~3, 4, and 5 at an intermediary node. The protocol is as follows (FIG.~\ref{fig:3}):

\begin{enumerate}
    \item The central nodes first locally prepare the $W_{\mathrm{mod}}$ states following the bipartite entropy distribution as mentioned above. 
    \item Once the preparation phases are over, both the central nodes distribute the locally prepared tripartite entangled states. The central node $C$ sends the qubits~ 1 and 2 to two end nodes and qubit~3 to an intermediary node $D$. In a similar way, the node $C^{\prime}$ sends qubit~6 to an end node and qubits~4 and 5 to the intermediary node.
    \item After receiving the qubits from the central nodes, the intermediary node performs a joint measurement on the received qubits using $W_{\mathrm{mod}}$-basis set.
    \item Based on the measurement outcome at the intermediary node, the end nodes apply local unitary corrections on their respective qubits.
\end{enumerate}

After all the steps mentioned above, a $W_{\mathrm{mod}}$ state is distributed across the end nodes among the qubits~1, 2, and 6. The detailed calculation corresponding to the protocol is provided in the Appendix \ref{appendix A2}.
%%%%%%%%%%%%%%%%%%%%%%%%%%%%%%%%%%%%%%%%%%%%%%%%%%%%%%%%%%%%%%%%%%%%%%%%
\section{Noise analysis for different distribution protocols}\label{sec4}
\subsection{Noise Model}

In realistic scenarios, qubits are not isolated from the environment. Therefore, inevitable interaction of the local qubits with their respective environment leads to decoherence, thereby degrading the global quantum correlations. To accurately characterize the decay of these correlations, we model the imperfections affecting each qubit as arising from \textit{depolarizing noise}, which is one of the most widely adopted and symmetry-preserving quantum noise models. The depolarizing channel provides a compact yet faithful representation of isotropic noise, encapsulating random bit-flip, phase-flip, and bit-phase-flip errors with equal likelihood. This choice ensures analytical tractability while capturing the essential physical processes that destroy entanglement in multi-qubit systems.

For a single qubit $\rho$, the depolarizing channel can be described using a completely positive trace-preserving (CPTP) map as
\begin{equation}
\mathcal{E}(\rho) = (1-p)\rho + \frac{p}{3}(X\rho X + Y\rho Y + Z\rho Z),
\end{equation}
where $X$, $Y$, and $Z$ denote the Pauli matrices, and $p \in [0,1]$ quantifies the strength of depolarization. Physically, the parameter $p$ represents the probability that the qubit undergoes a random Pauli error, while $1-p$ denotes the probability of leaving it unaffected.
An equivalent and operationally convenient Kraus representation is given by
\begin{equation}
\mathcal{E}(\rho) = \sum_{i=0}^{3} K_i \rho K_i^\dagger,
\end{equation}
where the Kraus operators are defined as
\begin{align}
K_0 &= \sqrt{1-\tfrac{3p}{4}}\,I, \nonumber\\
K_1 &= \sqrt{\tfrac{p}{4}}\,X, \nonumber\\
K_2 &= \sqrt{\tfrac{p}{4}}\,Y, \nonumber\\
K_3 &= \sqrt{\tfrac{p}{4}}\,Z.
\end{align}
with the condition $\sum_i K_i^\dagger K_i = I$, ensuring the map’s complete positivity and trace preservation. Under this noise process, the off-diagonal elements of $\rho$—which encode quantum coherence—decay linearly with $p$.

For a multiqubit state, assuming that every qubit undergoes the same depolarizing noise, the overall evolution of the state can be described by a tensor product of single-qubit depolarizing channels:
\begin{equation}
\mathcal{E}^{(n)} = \mathcal{E}_1 \otimes \mathcal{E}_2 \otimes \mathcal{E}_3 ... \otimes \mathcal{E}_n
\end{equation}
where the channels $\mathcal{E}_1, \mathcal{E}_2, \mathcal{E}_3, ~\text{and}~ \mathcal{E}_n$ are the ones acting on qubits 1, 2, 3, and $n$, respectively. If we consider a 3-qubit state $\rho_{123}$, then the action of the depolarizing channel can be expressed in terms of Kraus operators as
\begin{equation}
\rho'_{123}(p) = \sum_{i,j,k=0}^{3} K_{ijk}\, \rho_{123}\, K_{ijk}^\dagger,
\end{equation}
where each composite Kraus operator takes the form
\begin{equation}
K_{ijk} = K_i \otimes K_j \otimes K_k, \qquad i,j,k \in \{0,1,2,3\}.
\end{equation}
satisfying the normalization condition
\begin{equation}
\sum_{i,j,k} K_{ijk}^\dagger K_{ijk} = I_8.
\end{equation}

%%%%%%%%%%%%%%%%%%%%%%%%%%%%%%%%%
\begin{figure}[t]
    \centering
    % First figure: Fidelity vs p
    \begin{subfigure}[b]{0.5\textwidth}
        \centering
        \includegraphics[width=\textwidth]{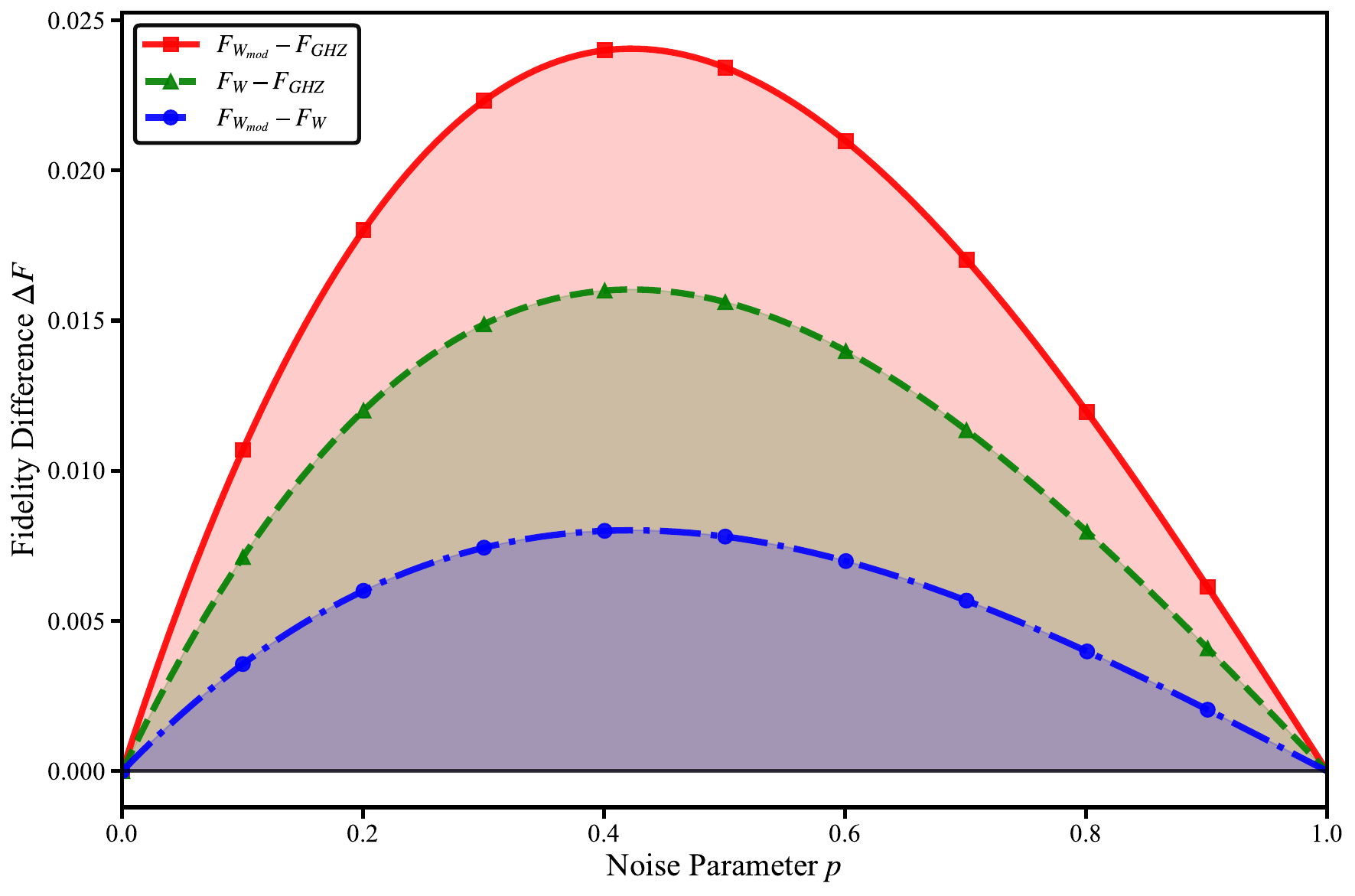}
        \caption{\small Variation of the fidelity difference in the presence of noise.}
        \label{fig:4a}
    \end{subfigure}
    \hfill
    % Second figure: Fidelity Difference
    \begin{subfigure}[b]{0.5\textwidth}
        \centering
        \includegraphics[width=\textwidth]{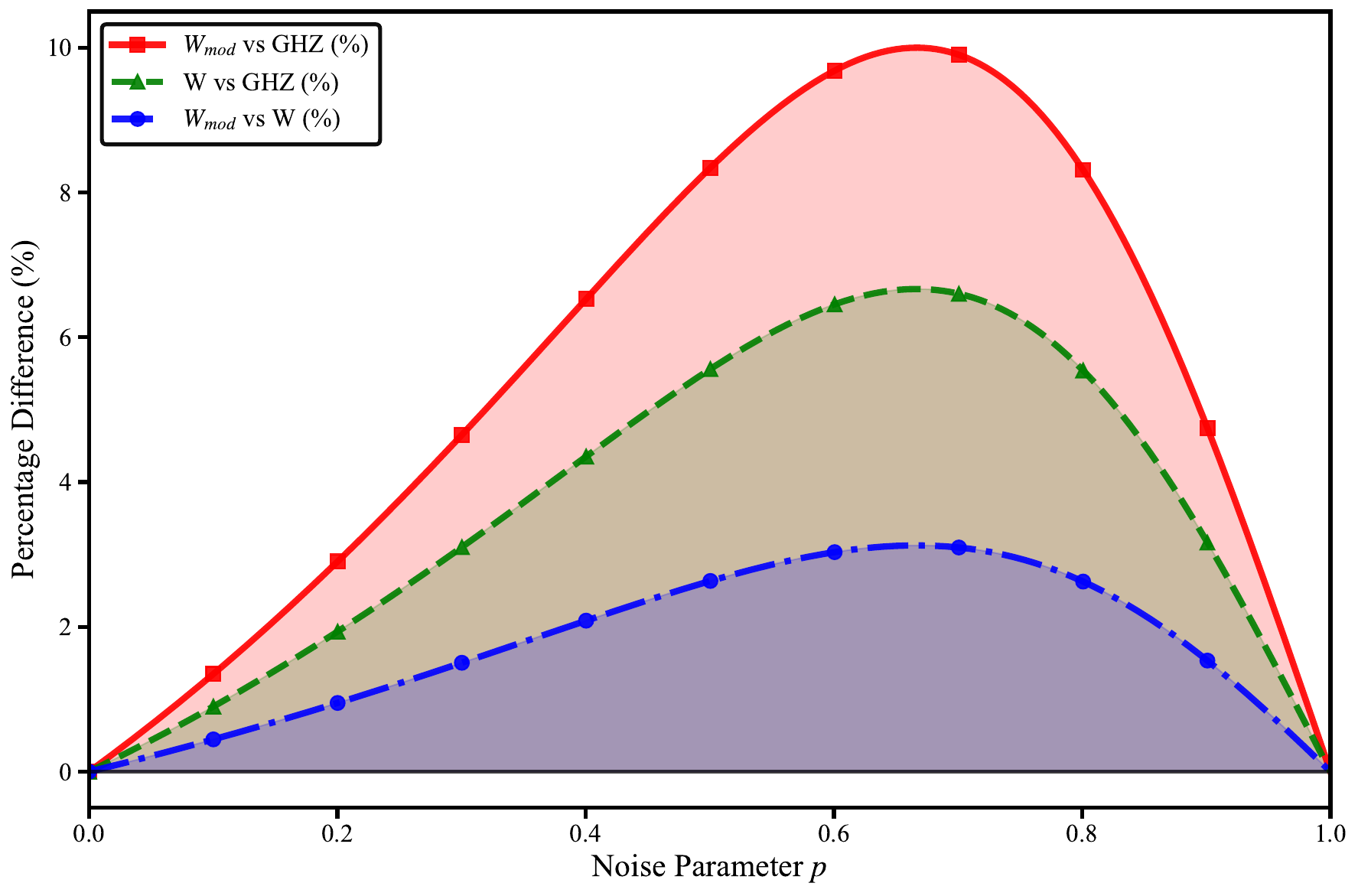}
        \caption{\small Variation of the percentage fidelity difference in the presence of noise.}
        \label{fig:4b}
    \end{subfigure}
    \caption{\justifying{Variation of a) fidelity difference b) percentage fidelity difference with respect to noise parameter $p$. The red, green, and blue lines represent the fidelity difference between $W_{\text{mod}}$ and $GHZ$, $W$ and $GHZ$, and $W_{\text{mod}}$ and $W$ states, respectively.}}
    \label{fig:4}
    \vspace{-2em}
\end{figure} 

The tensor-product structure of the individually acting Kraus operators ensures that errors occurring on different qubits are statistically uncorrelated—a realistic approximation for most network implementations where the qubits interact primarily with its local environments. This formalism provides a general and physically grounded framework for studying the decay of multipartite entanglement under isotropic, memoryless noise, serving as the foundation for our subsequent analysis of entanglement robustness and fidelity in $W_{\mathrm{mod}}$-state distribution protocols.
%%%%%%%%%%%%%%%%%%%%%%%%%%%%%%%%%%%%
\subsection{Performance analysis of the three distribution protocols }

To characterize the robustness of the protocols for distributing $W_{\mathrm{mod}}$ states in quantum networks, we analyze how bipartite and global quantum correlations degrade under the action of independent depolarizing noise on each qubit. We use tangles to characterize the entanglement of the $W$-class states~\cite{chaves}. For the $W$-class states three-tangle is zero \cite{three_tangle}, therefore two-tangles suffice to quantify the entanglement of the $W$-class states. We compute pairwise \textit{two-tangle} between the qubits and their average to characterizethe  \textit{global entanglement} of a $W$-class state. To calculate pairwise two-tangle, we first compute the concurrence $C_{ij}(p)$ for each qubit pair $\rho_{ij}(p)$ using the formula \cite{wootters1998}
\begin{equation}
C_{ij}(p) = \max\left\{0,\,\sqrt{\lambda_1(p)} - \sqrt{\lambda_2(p)} - \sqrt{\lambda_3(p)} - \sqrt{\lambda_4(p)}\right\},
\end{equation}
where \(\{\lambda_k(p)\}\) are the ordered eigenvalues of the matrix
\[
R_{ij}(p) = \rho_{ij}(p)\,(\sigma_y\otimes\sigma_y)\,\rho_{ij}(p)^*\,(\sigma_y\otimes\sigma_y).
\]
The corresponding two-tangle is defined as
\begin{equation}
\tau_{ij}(p) = C_{ij}^2(p).
\end{equation}

In the present study, all three pairwise two-tangles, \(\tau_{12}(p)\), \(\tau_{13}(p)\), and \(\tau_{23}(p)\),
are evaluated as functions of the depolarizing probability \(p\in[0,1]\). The critical noise thresholds \(p_c^{(ij)}\) are obtained as the smallest \(p\) for which \(\tau_{ij}(p)=0\), indicating complete bipartite separability.  
For symmetric states such as the ideal \(W\), all the three tangles are equal; for the $W_{\mathrm{mod}}$ state, asymmetry in amplitudes leads to distinct disentanglement rates across pairs.
To quantify the global entanglement in the distributed $W_{\mathrm{mod}}$ state, we employ average two-tangle.
For a state for a system `ABC', one can define average two-tangle as 
\begin{equation}\label{eq17}
\tau_{\text{av}}(p) = \frac{1}{3}\left[\tau_{AB}(p) + \tau_{BC}(p) + \tau_{CA}(p)\right].
\end{equation}

Here, {\it e.g.}, $\tau_{AB}(p)$ is two-tangle of the subsystem `AB'. Individual two-tangle gives us a local picture, while the average provides a global picture. When average is 
zero, the state is fully disentangled.

%we employ the Meyer--Wallach (MW) measure~\cite{meyer2002,brennen2003}, defined for a pure \(N\)-qubit state \(|\psi\rangle\) as
%\begin{equation}
%E_{\mathrm{MW}}(|\psi\rangle) = \frac{2}{N} \sum_{i=1}^{N}\left(1 - \mathrm{Tr}[\rho_i^2]\right),
%\end{equation}
%where \(\rho_i = \mathrm{Tr}_{\bar{i}}(|\psi\rangle\langle\psi|)\) is the single-qubit reduced state.  
%For a mixed state \(\rho'(p)\) obtained under noise, we extend this definition by computing \(E_{\mathrm{MW}}(\rho'(p))\) directly from the reduced single-qubit marginals. For three-qubit systems, the MW measure can be expressed in terms of the pairwise two-tangles~\cite{chaves} as:
%\begin{equation}
%E_{\mathrm{MW}}(p) = \frac{2}{3}\left[\tau_{12}(p) + \tau_{13}(p) + \tau_{23}(p)\right].
%\label{eq:mw_sum_tangles}
%\end{equation}

\subsubsection{Protocol 1}
To find out the robustness of the various distribution protocols in noisy scenarios, we use two performance metrics -- \textit{fidelity} of the distributed states with respect to the target reference state and \textit{entanglement decay} of the distributed states. We first compute the fidelity of the $W_{\mathrm{mod}}$ states as a function of the depolarizing parameter $p$ using the formula,
\begin{equation}
F(p)\;=\;\langle\psi|\rho'(p)|\psi\rangle.
\end{equation}
Assuming that in this protocol identical and independent depolarizing noise acts on each qubit, we find the fidelity to be
\begin{equation}
F_{W_{\mathrm{mod}}}(p)=1-\tfrac{17}{8}p+\tfrac{27}{16}p^2-\tfrac{7}{16}p^3.
\end{equation}
From this expression it is clear that when there is no noise ($p=0$) the value of fidelity is unity, whereas, when the noise is maximum ($p=1$), fidelity becomes 1/8, implying that the resultant state is a maximally mixed state of three qubits. When compared with the fidelities of GHZ and W states, distributed using this protocol, we find that this protocol is more robust for distributing $W_{\mathrm{mod}}$ states.

FIG.~\ref{fig:4} shows how the fidelity of the GHZ, $W$, and $W_{\mathrm{mod}}$ states as a function of the depolarizing probability $p$. All states exhibit a monotonic decay in fidelity with increasing noise strength, consistent with the uniform loss of coherence induced by the isotropic depolarizing channel. From the graphical plots it is clear that the decay of fidelity is fastest in the case of GHZ states, while both $W$-class states retain higher values throughout the entire range of $p$. Notably, the $W_{\mathrm{mod}}$ state consistently outperforms the standard $W$ state, indicating an enhanced robustness arising from its asymmetric amplitude distribution. The corresponding fidelity-difference plot further reconfirms the trend. The gap between $W_{\mathrm{mod}}$ and $W$ remains comparatively small but positive throughout, demonstrating that even a moderate redistribution of excitation amplitudes improves noise tolerance. These observations confirm that the $W_{\mathrm{mod}}$ states are more resilient to depolarizing noise, making it a more suitable candidate for direct entanglement distribution in noisy quantum networks.

\paragraph*{}

 \begin{figure}[t]
    \centering
    \includegraphics[width=1\linewidth]{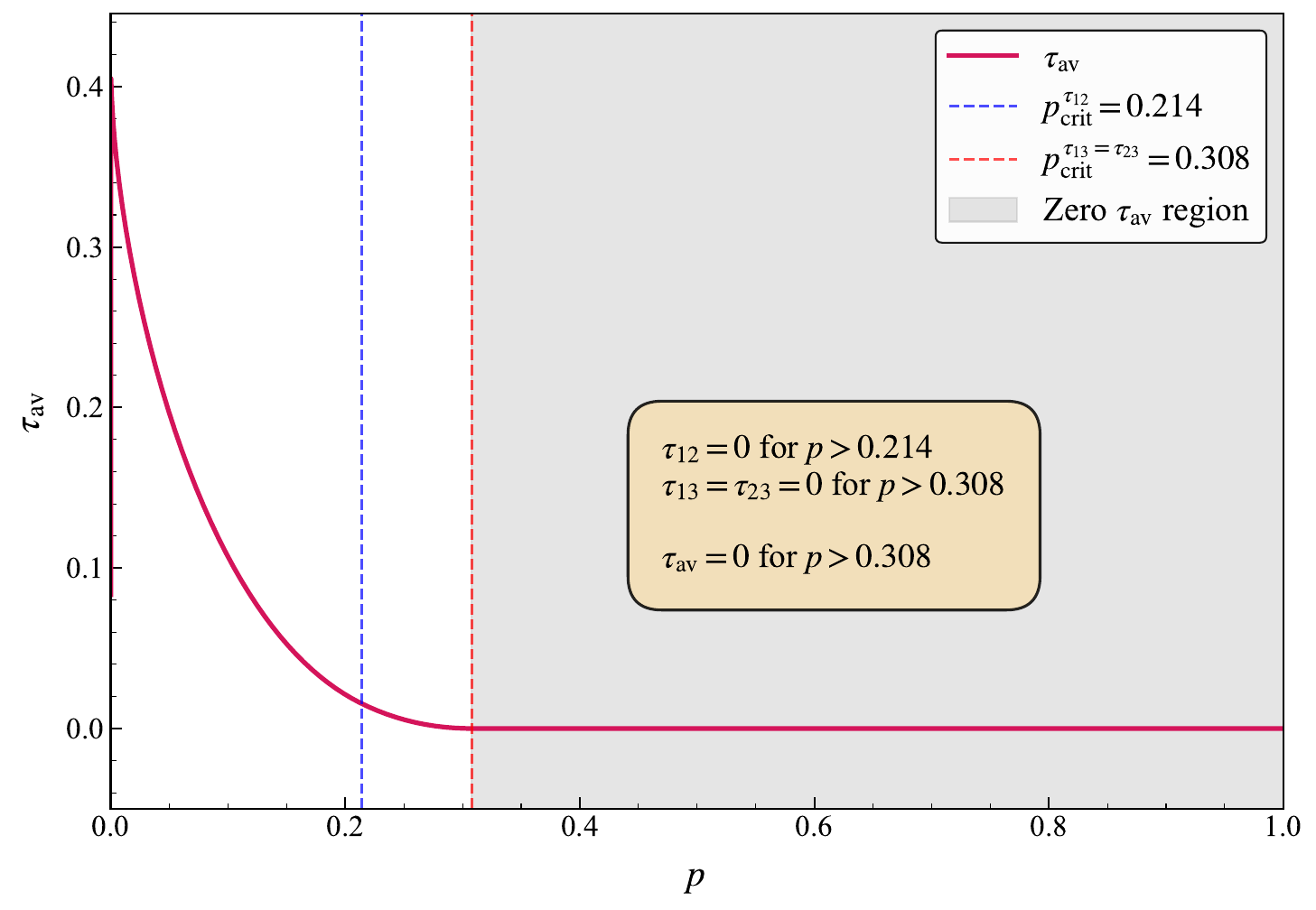}
    \caption{\justifying{Decay of global entanglement of distributed $W_{\mathrm{mod}}$ state when the qubits are sent directly to the end nodes from the central node. The vertical dotted blue and red lines enclose the regions in which tangle between qubits 1 and 2 and tangle between qubits 3 and other two are non-zero, respectively. The white and gray regions represent non-zero and zero global entanglement, respectively.}}
    \label{fig:5}
\end{figure}
The explicit relation (Eq. {\ref{eq17}}) between the global entanglement and pairwise two-tangle helps in determining the change in the former quantity with respect to the decay in the later quantities. The noise threshold \(p_c^{(G)}\) corresponding to complete loss of global entanglement is determined as the smallest \(p\) such that $\tau_{\text{av}}(p)=0$, which typically occurs when the last surviving two-tangle vanishes (FIG.~{\ref{fig:5}}).
The simultaneous analysis of qubit-pairwise and global entanglement provides a clear operational interpretation of the performance of this distribution protocol. The points \(p_c^{(12)}, p_c^{(13)}, p_c^{(23)}\) mark the noise thresholds beyond which entanglement between the node pairs completely decays, leading to loss of two-party quantum communication. The retention  of a finite $\tau_{\text{av}}(p)$ below these thresholds indicates the existence of global quantum correlations in the network, which can support multipartite tasks such as joint teleportation or distributed state reconstruction.

%%%%%%%%%%%%%%%%%%%%%%%%%%%%%
\subsubsection{Protocol 2}
In this protocol, the initially prepared $W_{\mathrm{mod}}$ state,
\begin{equation}
\ket{W_{\mathrm{mod}}}_{123}=\frac{1}{2}\ket{100}_{123}+\frac{1}{2}\ket{010}_{12 3}+\frac{1}{\sqrt{2}}\ket{001}_{123}\nonumber\\
\end{equation}
is distributed across the quantum network by sending qubits~ 1 and 2 directly to two end nodes and injecting qubit~3 into a linear repeater chain using Bell pairs and BSMs. We consider that the linear repeater chain has $n-1$ number of repeaters, which helps in sending qubit~3 from the central node to the final end node after $n$ hops. Each repeater has two memories involved, and the central and final end node hold one memory each. The BSMs are conducted in the central node and the repeater nodes. Considering this scenario, we now find out the noise acting on each qubit during the distribution of the target state. 
For qubits~ 1 and 2, we consider that the noise arises only due to transmission links between them and the central node, and the noise acts once. Let $p$ be the depolarising noise parameter for these qubits. However, qubit~3 will suffer from noise arising from three sources -- transmission links in between the central and the final node, memories, and BSMs, and these various noise sources will act multiple times on the qubit. If we consider that the noise arising from various sources are of the same nature, i.e, $p_{\text{link}} = p_{\text{mem}} = p_{\text{BSM}} = p$, where $p_{\text{link}},~ p_{\text{mem}}, ~\text{and}~p_{\text{BSM}}$ are noise parameters corresponding to links, memories, and BSMs,  then it is quiet straightforward to find the total number of such depolarising noise events. We find,
\begin{equation}
E_{\text{tot}} = E_{\text{link}} + E_{\text{mem}} + E_{\text{BSM}}
= n + (2n -1) + n = 4n - 1.
\end{equation}
where, $E_{\text{tot}},~E_{\text{link}},~E_{\text{mem}},~\text{and}~E_{\text{BSM}}$ are the total depolarising noise events and noise events corresponding to links, memories, and BSMs, respectively. Now, as the composition of depolarising channels is also a depolarising channel, the effective noise parameter on qubit~3 will be a single depolarising channel with parameter $p_{\text{eff}}$, which can be explicitly expressed as,

\begin{eqnarray}
1 - p_{\text{eff}}(n)&=& (1 - p_{\text{link}})^n (1-p_{\text{mem}})^{2n-1} (1-p_{\text{BSM}})^n\nonumber\\
p_{\text{eff}}(n) &=& 1 - (1-p)^{4n-1}.
\end{eqnarray}

\begin{figure}[t]
    \centering
    \begin{subfigure}{0.5\textwidth}
    \centering
    \includegraphics[width=\linewidth]{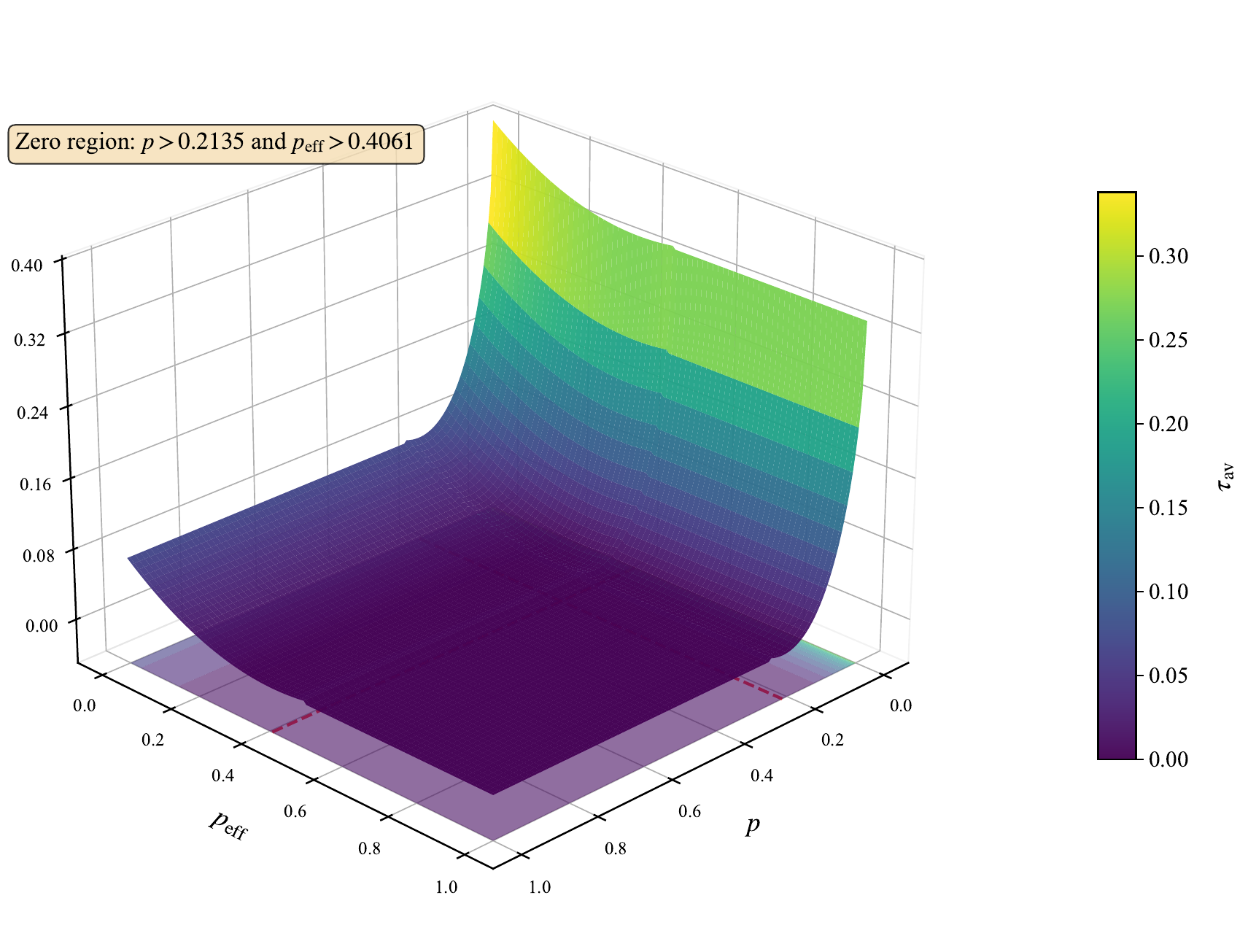}
    \caption{Evolution of total entanglement with respect to the noise parameters $p$ and $p_{\text{eff}}$.  }
    \label{fig:6a}
    \end{subfigure}
    \begin{subfigure}{.5\textwidth}
  \centering
    \includegraphics[width=1\linewidth]{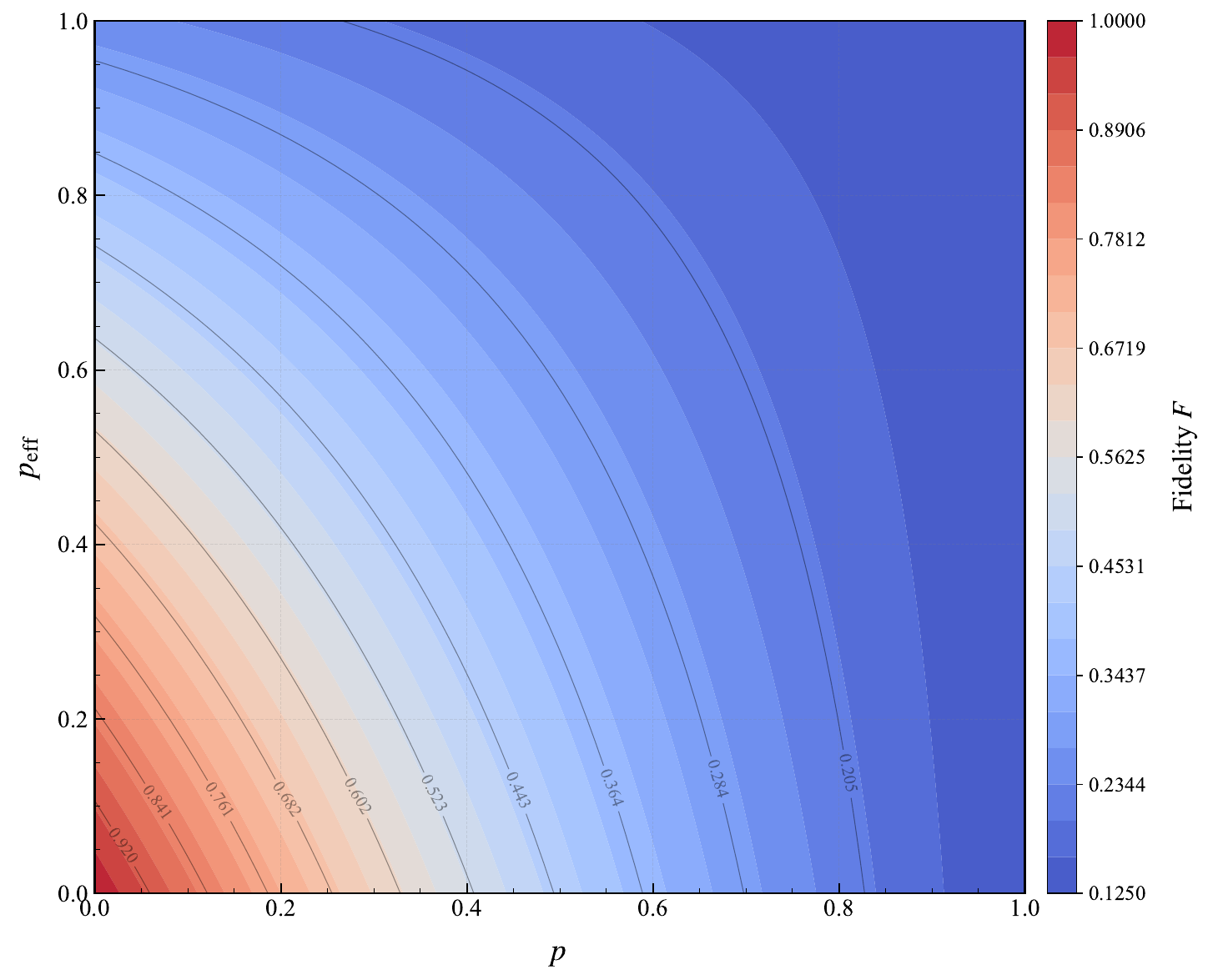}
    \caption{\justifying{Fidelity change of the distributed $W_{\mathrm{mod}}$ state with noise for second protocol.}}
    \label{fig:6b}
     \end{subfigure}
     \caption{\justifying{Evolution of a) total entanglement b) fidelity change of the $W_{\mathrm{mod}}$ state that is distributed across end nodes by sending qubits 1 and 2 directly and qubit 3 through the repeater network.}}
     \label{fig:6}
\end{figure}

Hence, the resultant noisy $W_{\mathrm{mod}}$ state $\rho_{125}$ at the end of the distribution protocol after $n$ hops can be derived using Eq. (11). Now we analyze the robustness of this protocol using fidelity and entanglement dynamics of the final state. As the state is a function of noise parameters $p$ and $p_{\text{eff}}$, fidelity is also a function of the same parameters. To find out how the fidelity of the final state varies with the noise parameters, we plot a surface graph, as shown in FIG. \ref{fig:6a}. From the graph, it is clear that as the depolarisation noise parameters increase, the final distributed state no longer remains entangled. To study the effect of noise on this distribution protocol, we further analyze the entanglement dynamics of the final state $\rho_{125}$. First of all, we find the value of total entanglement $\tau_{av}$ of the state by evaluating the values of $\tau_{12}, \tau_{15},~ \text{and}~ \tau_{25}$. 

The total entanglement is a function of both the noise parameters, and when we plot it against noise parameters (FIG.~{\ref{fig:6a}), we get threshold values of the parameters for which the value of total entanglement is non-zero. We find that for $p=0.214~ \text{and}~ p_{\text{eff}} = 0.408$, all the tangles as well as the total entanglement are non-zero. For $p$-values greater than 0.214, the tangle between qubits~1 and 2 becomes zero, however, total entanglement remains non-zero. When the depolarising noise becomes large enough for the values of the parameters to go beyond the threshold, all the tangles and total entanglement vanish. 

Now, to find out how far qubit~3 can be sent through entanglement swapping, it is necessary to calculate the maximum number of hops and total length between the central and end node for any particular value of $p_{\text{eff}}$ below threshold. For doing so, we express $p_{\text{eff}}$ explicitly as a function of $n$ and total length $L$. We first relate the single-event error $p$ to the link length $l$ via an exponential noise model, $p = 1 - e^{-\lambda l}$, where $\lambda$ is an effective attenuation rate and $l$ is the length of each elementary link. Substituting this value of $p$ in Eq. (21) we get,
\begin{equation}
p_{\text{eff}} = 1 - \big(e^{-\lambda l}\big)^{4n - 1}.
\end{equation}
Expressing the total distance as $L = n l$, we get the effective depolarising parameter as,
\begin{equation}
p_{\text{eff}}(n) = 1 - e^{-(4n - 1)\lambda L / n}.
\end{equation}

FIG.~\ref{fig:7} presents a comprehensive analysis of the quantum 
repeater network design space under realistic noise constraints. The parameter space 
map (FIG.~\ref{fig:7a}) visualizes the effective noise parameter 
$p_{\mathrm{eff}}(n,\ell) = 1 - \exp[-(4n-1)\lambda \ell]$ across different 
network configurations, where $n$ denotes the number of repeater stations, $\ell$ 
represents the elementary link length, and $\lambda = 0.046~\mathrm{km}^{-1}$ is 
the fiber attenuation parameter. The color gradient transitions from green (low noise) 
to red (high noise), with the green shaded region indicating feasible configurations 
that satisfy the noise threshold $p_{\mathrm{eff}} \leq 0.407$. The solid green 
boundary curve demarcates the maximum allowable link length for each repeater number, 
beyond which the accumulated noise exceeds the acceptable limit. Black contour lines 
at $p_{\mathrm{eff}} = 0.2$, $0.3$, and $0.407$ (bold) provide quantitative 
reference points, revealing the rapid contraction of the feasible design space as 
$n$ increases. We highlight four selected operating configurations with red stars: 
$(n=1, \ell=3.5~\mathrm{km}, p_{\mathrm{eff}}=0.383)$, 
$(n=2, \ell=1.5~\mathrm{km}, p_{\mathrm{eff}}=0.383)$, 
$(n=3, \ell=1.0~\mathrm{km}, p_{\mathrm{eff}}=0.397)$, and 
$(n=10, \ell=0.29~\mathrm{km}, p_{\mathrm{eff}}=0.407)$. These configurations 
demonstrate the fundamental trade-off: while increasing $n$ allows more 
entanglement swapping operations, it also necessitates dramatically shorter 
link lengths due to the $(4n-1)$ noisy operations per protocol execution.

The quantitative data in FIG.~\ref{fig:7b} reinforces this observation, 
tabulating maximum allowable link lengths $\ell_{\max}$ and total network distances 
$L_{\max} = n\ell_{\max}$ for integer repeater numbers at three noise thresholds. 
The constraint $(4n-1)\ell_{\max} = -\ln(1-p_{\mathrm{eff}})/\lambda$ directly 
determines the boundary, causing $\ell_{\max}$ to decrease as $1/(4n-1)$ while the 
total distance $L_{\max}$ asymptotically approaches a limiting value. For instance, 
at $p_{\mathrm{eff}} = 0.407$, increasing from $n=1$ to $n=20$ reduces $\ell_{\max}$ 
from $3.79~\mathrm{km}$ to $0.14~\mathrm{km}$---a 26-fold decrease---while $L_{\max}$ 
only decreases from $3.79~\mathrm{km}$ to $2.88~\mathrm{km}$, showing diminishing 
returns beyond approximately $n=5$ repeater stations. 

%%%%%%%%%%%%%%%%%%%%%%%%%%%%%
\subsubsection{Protocol 3}

In this protocol, two central nodes $C$ and $C^{\prime}$ share a $W_{\mathrm{mod}}$ state between three end nodes through a joint 3-qubit measurement at an intermediate node and direct transmission of qubits to the end nodes. The qubits at the end nodes are projected onto a $W_{\mathrm{mod}}$ state after the joint-measurement is performed at the intermediate node. If the qubits on which the measurement is performed do not suffer from noise, then the final state formed by the other qubits is a pure $W_{\mathrm{mod}}$ state. However, if those qubits are subjected to noise, then the resultant state will be a noisy one. Therefore, robustness of this protocol depends on how noise affects the qubits at the intermediate node. Noise sources affecting the qubits at the intermediate node are --  links through which
the qubits are sent from the central nodes to the intermediate node, memories at the intermediate node in which the received qubits are stored, and the noisy joint-measurement performed on those qubits. Just like the second protocol, we consider the effective noise to be depolarising with the parameter $p_{\text{eff}}$. After the joint measurement and tracing out of the qubits on which the measurement is carried on, this effective noise passes on to the qubits at the end nodes, resulting to a noisy $W_{\mathrm{mod}}$ state. It can be shown that this whole process is equivalent to applying the effective noise to a pure $W_{\mathrm{mod}}$ state formed by the qubits at the end nodes. Therefore, the final noisy state that we get can be expressed as,
\begin{equation}
\rho_{126} = \beta^3|W_{\mathrm{mod}}\rangle\ _{126}\langle W_{\mathrm{mod}}|_{126} + \frac{1 - \beta^3}{8}I_8,
\end{equation}
where $\beta = 1-\frac{3p_{\text{eff}}}{4}$ and $I_8$ is the 8 $\times$ 8 identity matrix. The effect of noise on the distributed entangled state can be analyzed if we calculate the tangles $\tau_{12}, \tau_{26},\tau_{16}$ and the total entanglement $\tau_{av}$ of the state given above. We find that the two-tangle between the qubits 1 and 2 first decreases with increasing $p$, reaches a minimum at $p = 0.392$, and again increases (FIG.~\ref{fig:8}). Whereas, $\tau_{16}~(\tau_{26})$ monotonically decreases with $p$ (FIG.~\ref{fig:9}). As the noise is not directly acting on the qubits 1 and 2, these two qubits never get completely decohered. When we analyze the evolution of total entanglement of the final distributed $W_{\text{mod}}$state, we find that the global entanglement is not completely destroyed due to noise, rather, it reaches a minimum value for $p = 0.392$ and then remains constant with increasing $p$ ((FIG.~\ref{fig:10})). 
%%%%%%%%%%%%%%%%%%%%%%
\begin{figure}[h]
    \centering
    \begin{subfigure}{.55\textwidth}
    \centering
    \includegraphics[width=\linewidth]{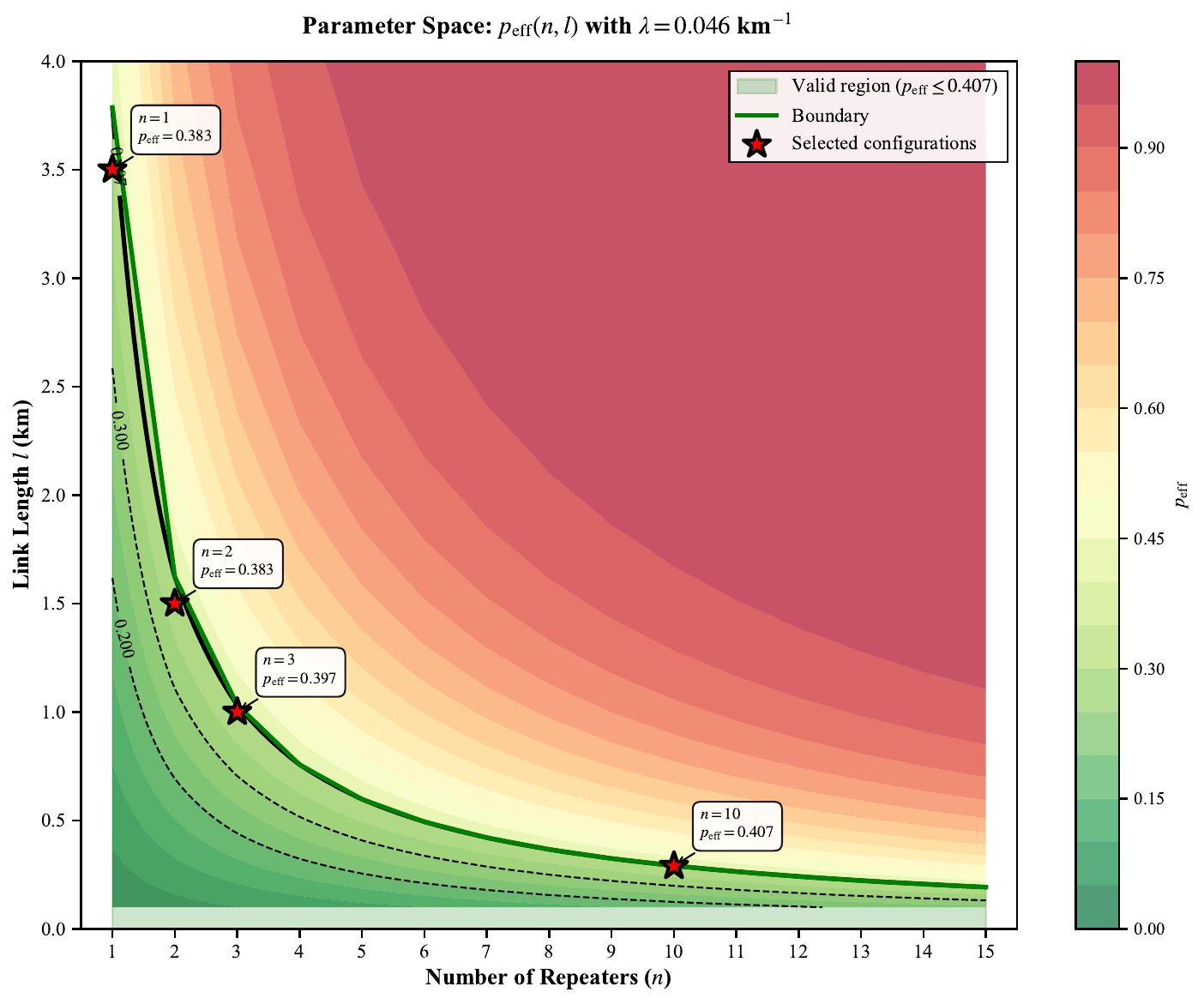}
    \caption{\justifying{}}
    \label{fig:7a}
\end{subfigure}
\begin{subfigure}{.55\textwidth}
    \centering
    \includegraphics[width=1\linewidth]{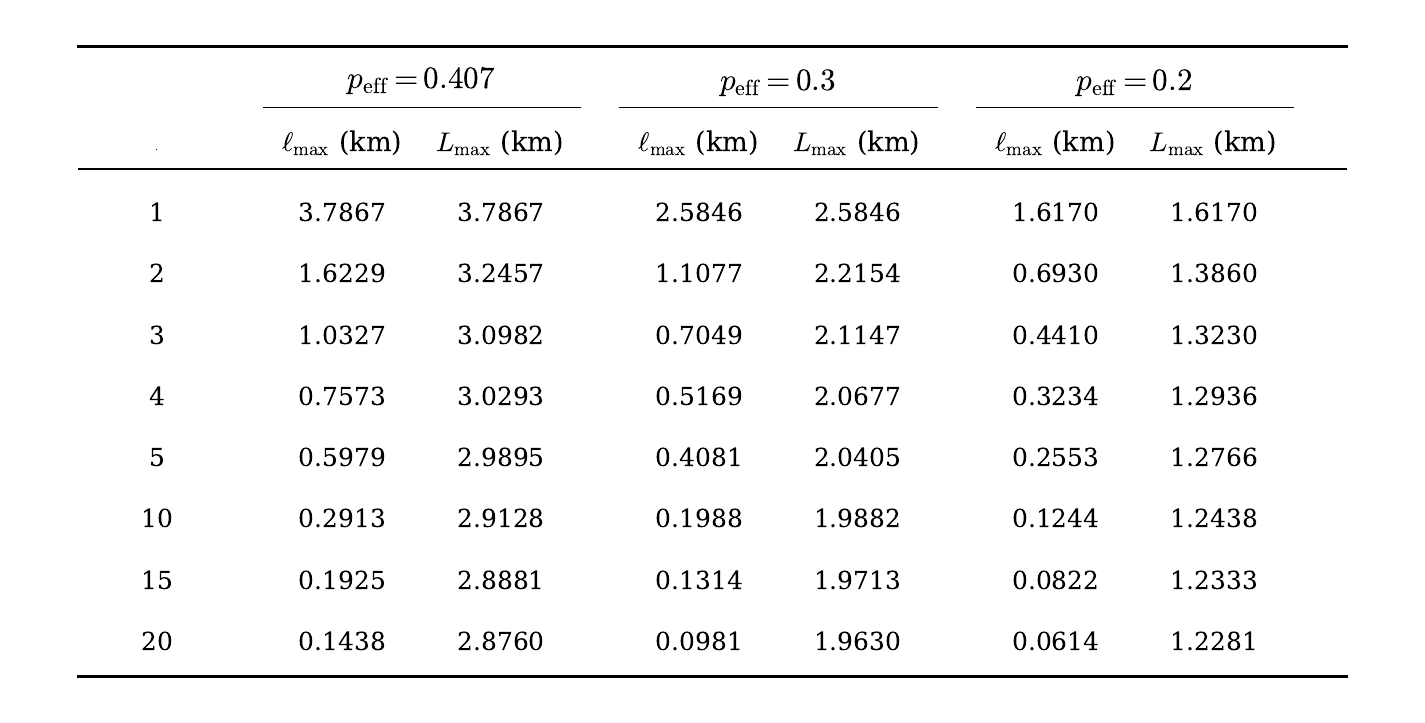}
    \caption{\justifying{}}
    \label{fig:7b}
\end{subfigure} 
\caption{\justifying{Quantum network parameter space and design constraints. 
(a) Color map of $p_{\mathrm{eff}}(n,\ell)$ showing feasible 
region (green, $p_{\mathrm{eff}} \leq 0.407$) and selected configurations 
(red stars). (b) Numerical values of $\ell_{\max}$ and $L_{\max}$ 
for $p_{\mathrm{eff}} = 0.2$, $0.3$, and $0.407$. The analysis reveals that 
adding more repeaters does not extend the maximum reachable distance under fixed 
noise constraints; instead, the optimal strategy depends on the specific operational 
requirements and hardware limitations of the quantum network implementation.}}
     \label{fig:7}
\end{figure}

%%%%%%%%%%%%%%%%%%%%%%%%%%%%%%%%%%%%%%%%%%%%%%%%%%%%%%%%%%%%%%%%%%%%%%
\section{Robustness of $W_{\mathrm{mod}}$ states in terms of coefficients}\label{sec5}
In the previous section, we have studied the robustness of various distribution protocols by analyzing how fidelity and entanglement of the final distributed state vary with noise parameters. In all of our analysis, we have considered the $W_{\mathrm{mod}}$ state with $m = 1$. However, if we consider the general $W_{\mathrm{mod}}$ state given in Eq.(\ref{eq0}), then fidelity and global entanglement of the final state will be a function of noise parameter and $m$.Here we investigate how robustness of $W_{\mathrm{mod}}$ states changes with increasing $m$ in a noisy scenario by considering protocol 1, where identical independent depolarising noise acts on each qubits. From our analysis, we find that as $m$ is increased, the fidelity of the corresponding W state increases compared to that of the state we have considered earlier, though the nature of variation of fidelity remains the same. However, the increment of fidelity becomes saturated as $m$ becomes sufficiently large, which in our case is around $m = 150$.In FIG. ~\ref{fig:11}, we present our analysis. 

\begin{figure}[t]
    \centering
    \includegraphics[width=1.05\linewidth]{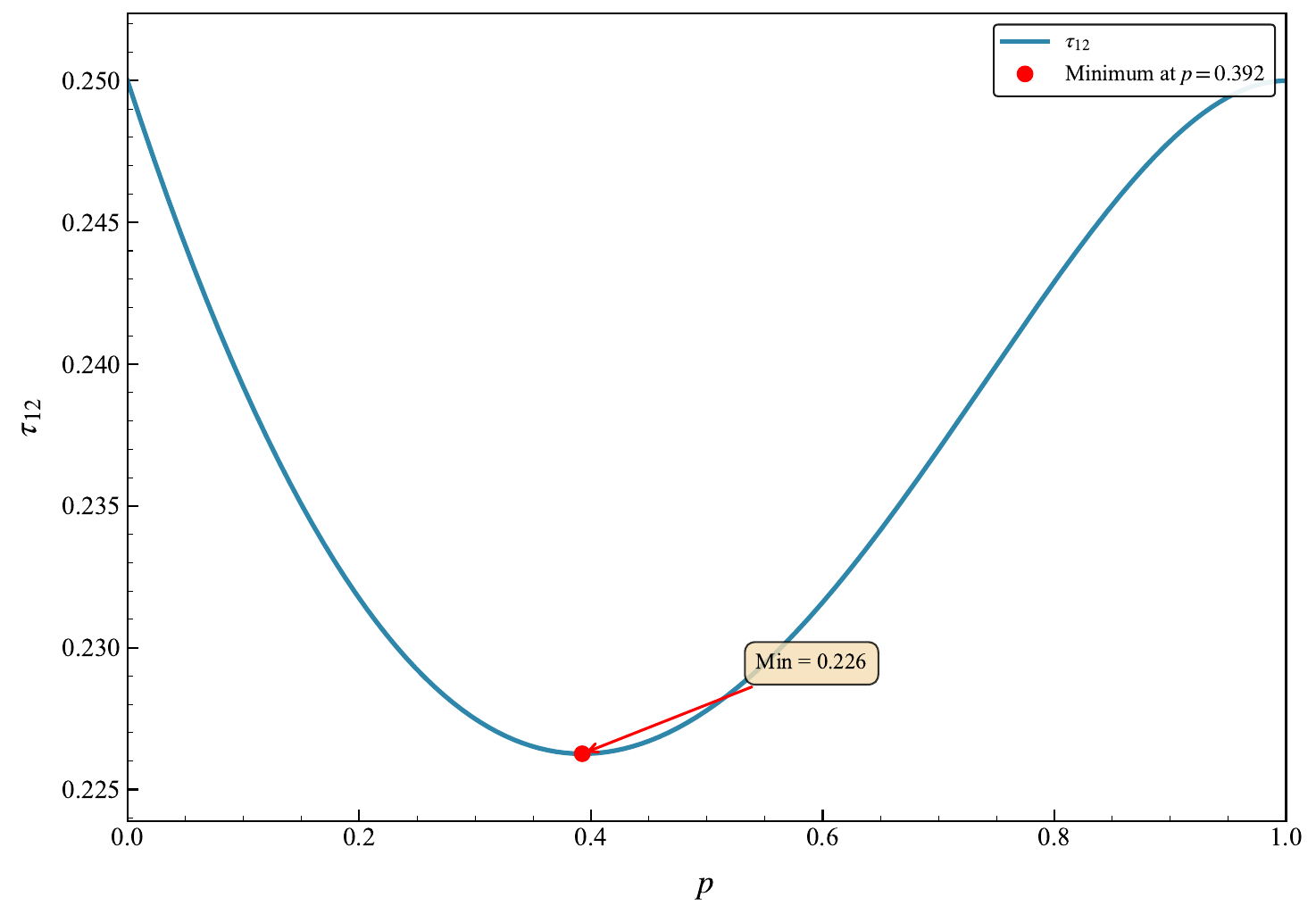}
    \caption{\justifying{The plot shows how two-tangle between qubits 1 and 2 evolves with respect to the noise parameter $p$ in the third protocol. Here $\tau_{12}$ never becomes zero with the increase of the noise parameter. }}
    \label{fig:8}
\end{figure}

\begin{figure}[t]
    \centering
    \includegraphics[width=1.05\linewidth]{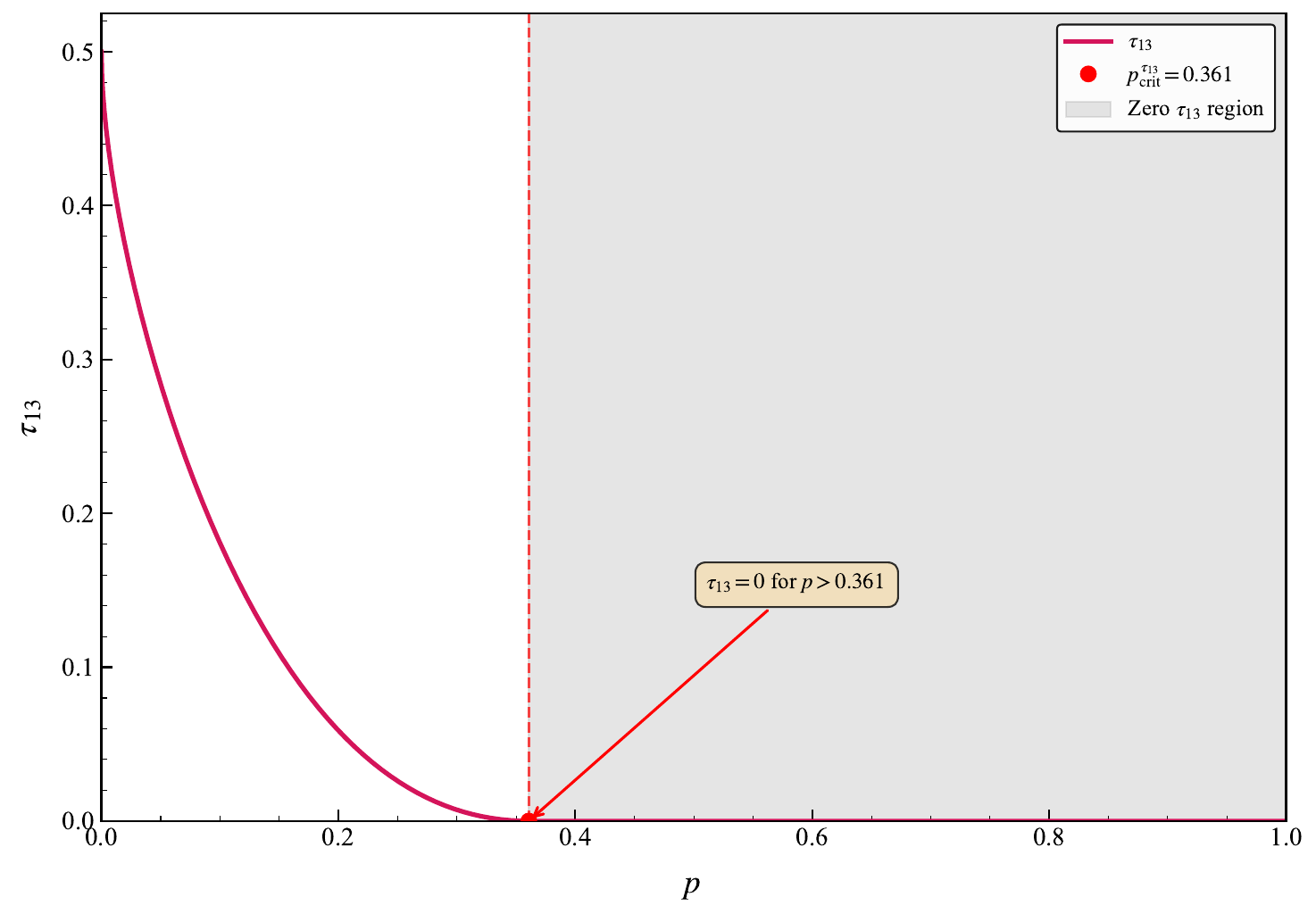}
    \caption{\justifying{Change of tangle between qubits 1 (2) and 6} with respect to the noise parameter $p$ in the third protocol. Unlike $\tau_{12}$, two-tangle $\tau_{16}$ reaches a minimum of zero for $p>0.361$.}
    \label{fig:9}
\end{figure}

\begin{figure}[t]
    \centering
    \includegraphics[width=1.05\linewidth]{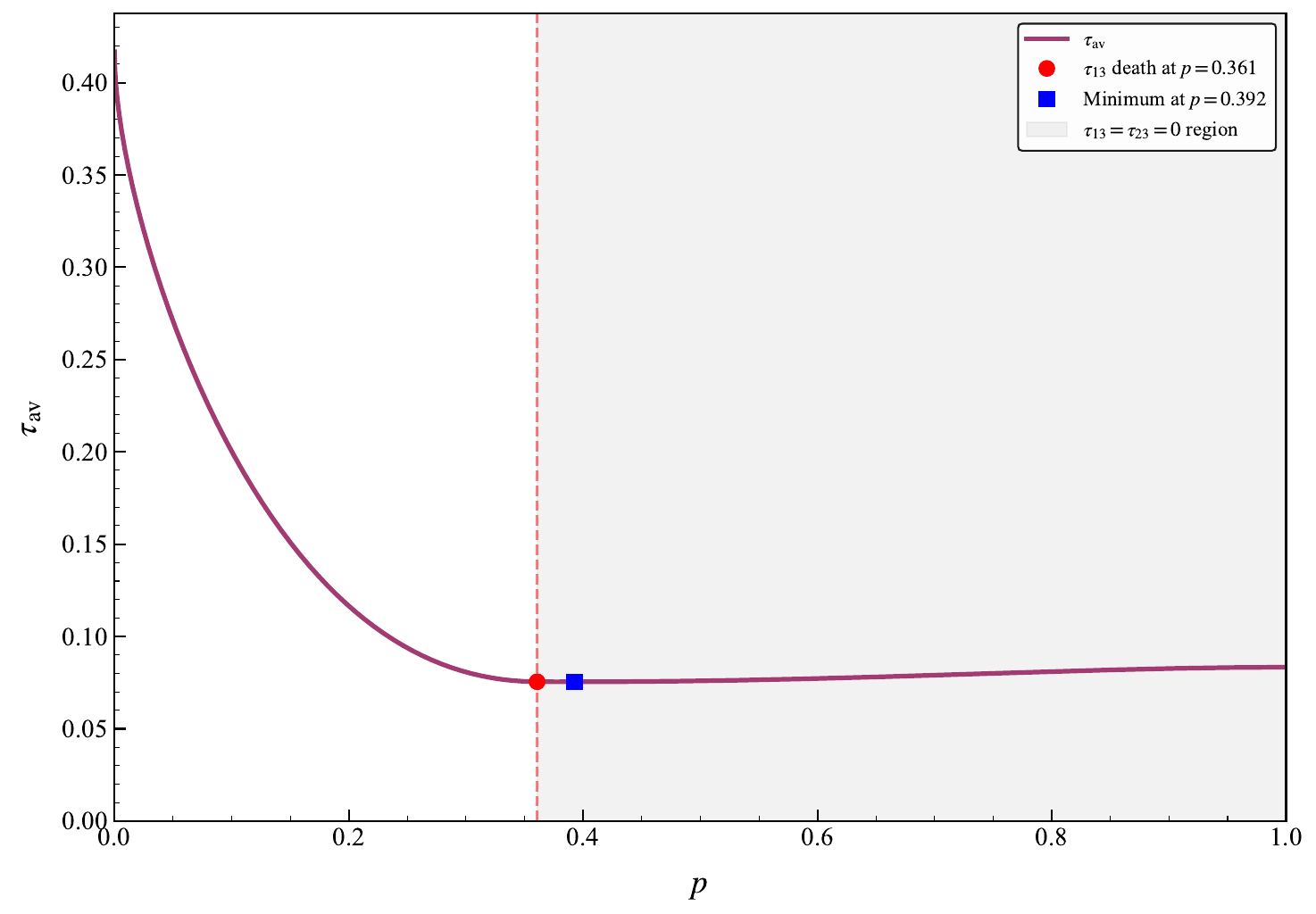}
    \caption{\justifying{Decay of global entanglement of the distributed $W_{\text{mod}}$ with noise in the third protocol. The global entanglement does not decay completely with increasing $p$ due to the non-vanishing $\tau_{12}$. In the white region all the two-tangles are non-zero, whereas, in the gray region only $\tau_{12}$ is non-zero.}}
    \label{fig:10}
\end{figure}

The saturation of fidelity observed around $m \approx 150$ can be physically attributed to the asymptotic behavior of the state amplitudes. In the limit $m \to \infty$,the coefficient of the $|100\rangle$ term vanishes, causing the tripartite system to effectively decouple. The state asymptotically approaches the product form $|0\rangle_1 \otimes |\Psi^+\rangle_{23}$, where $|\Psi^+\rangle_{23}$ represents a 
maximally entangled Bell state shared between qubits 2 and 3. Consequently, the noise robustness saturates at the intrinsic limit of bipartite entanglement, offering no further advantage beyond this threshold.

\begin{figure}[!htbp]
    \centering
    % First figure: Fidelity vs p
    \begin{subfigure}[b]{0.48\textwidth}
        \centering
        \includegraphics[width=\textwidth]{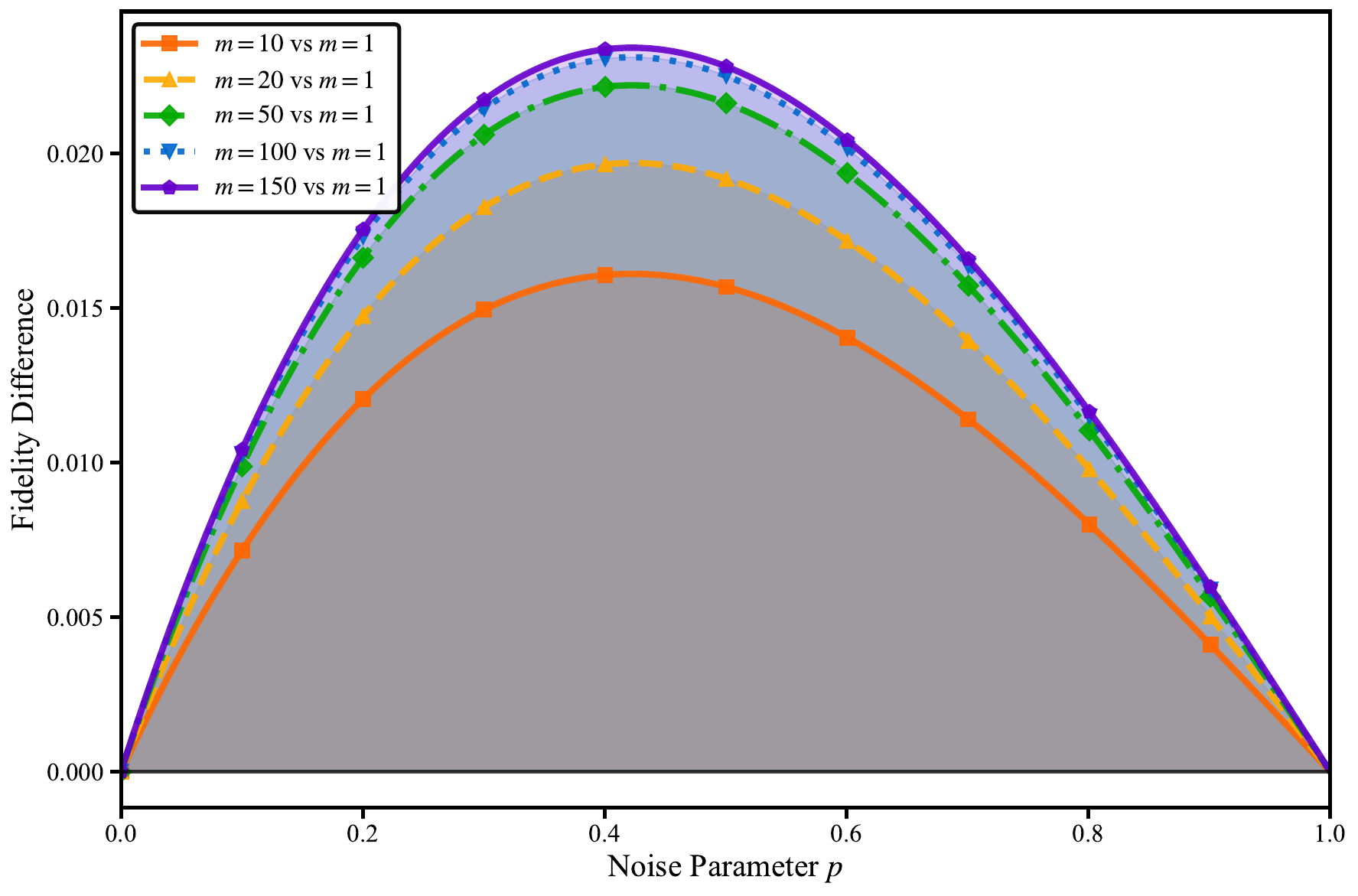}
        \caption{\small Variation of fidelity difference in the presence of noise.}
        \label{fig:11a}
    \end{subfigure}
    \hfill
    % Second figure: Fidelity Difference
    \begin{subfigure}[b]{0.48\textwidth}
        \centering
        \includegraphics[width=\textwidth]{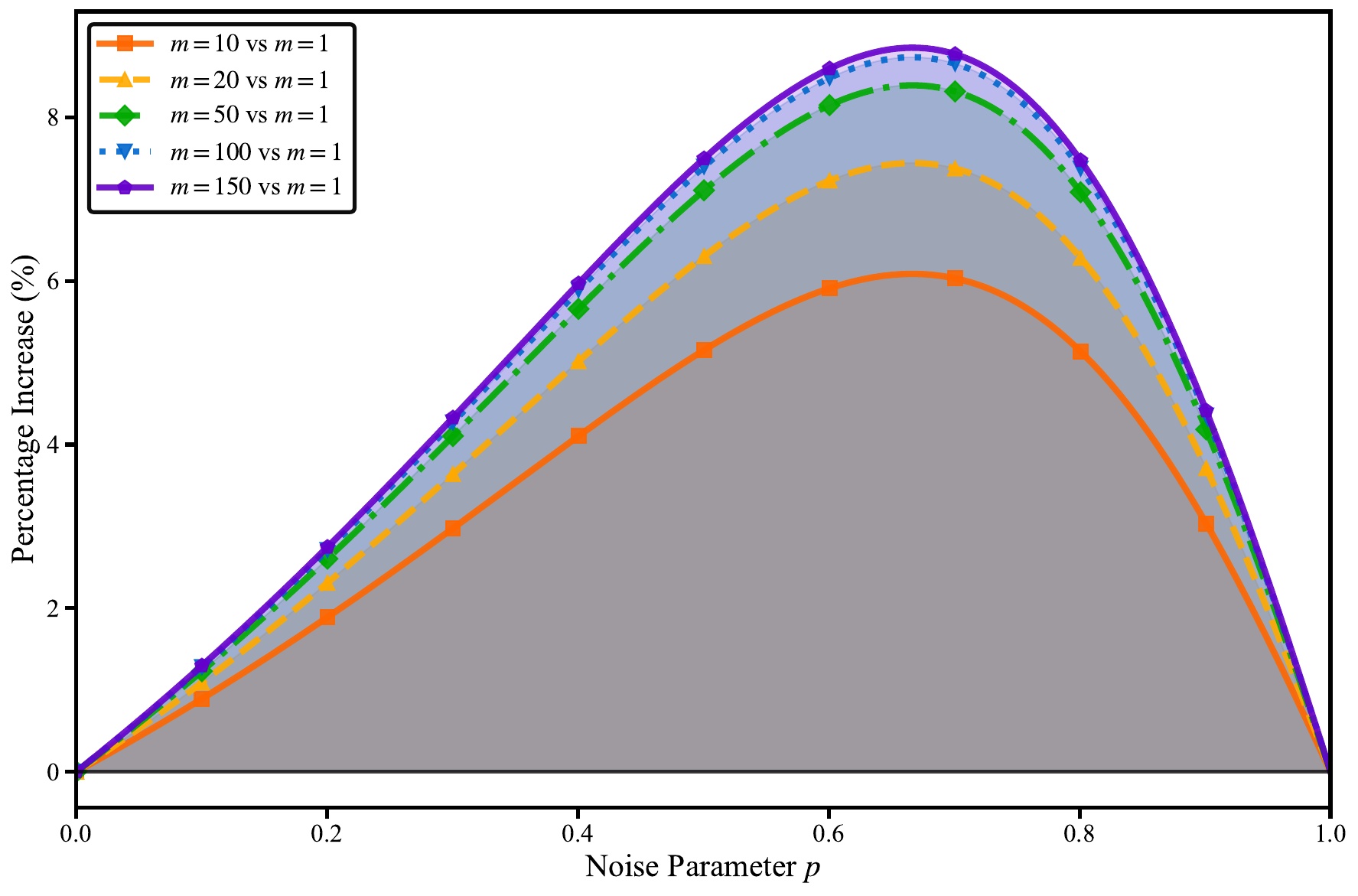}
        \caption{\small Variation of percentage fidelity difference in the presence of noise.}
        \label{fig:11b}
    \end{subfigure}
    \caption{\justifying{Plots showing increment of fidelity with increasing $m$ for a general $W_{\mathrm{mod}}$ state as a function of the noise parameter $p$. Identical independent depolarising noise with parameter $p$ is acting on each of the qubits. At $m = 150$, the increment in fidelity becomes saturated, implying that further increasing $m$ will not provide any advantage compared to the specific state that we have considered.}}
    \label{fig:11}
\end{figure}

%%%%%%%%%%%%%%%%%%%%%%%%%%%%%%%%%%%%%%%%%%%%%%%%%%%%%%%%%%%%%%%%%%%%%
\section{Conclusion and Discussion}\label{sec6}
Compared to GHZ states, $W_{\mathrm{mod}}$ states offer dual advantages -- these states can be used as a resource for deterministic quantum information-processing tasks and these states are robust against qubit loss. In this work, we have studied the distribution of $W_{\mathrm{mod}}$ states in quantum networks considering both ideal and noisy scenarios. In previous work the advantage of using central nodes over direct transmission considering the distribution of GHZ states was demonstrated. Here we describe three protocols for distributing $W_{\mathrm{mod}}$ states in quantum networks having central-node architecture. In the first protocol, the qubits are directly transmitted to the end nodes, whereas, in the two other protocols these are sent indirectly through BSM and multi-qubit joint measurement. To analyze the robustness of the protocols in the presence of noise, we study how fidelity and entanglement of the distributed state vary with the noise parameters during each protocol. We find that in the first protocol, where independent depolarising noise with parameter $p$ acts on the three qubits of the $W_{\mathrm{mod}}$ state, the global entanglement $\tau_{av}$ remains non-zero for $p< 0.308$. While $\tau_{12}$ between qubits 1 and 2 vanishes beyond a threshold of $p = 0.214$, qubit 3 remains entangled with the other qubits below the noise threshold $p = 0.308$, i.e., $\tau_{13}~\text{and}~\tau_{23}$ are non-zero even when $\tau_{12}$ becomes zero. What we find interesting is that in the second protocol, the noise threshold for $\tau_{12}$ to be zero is the same as that of first protocol, whereas, the threshold for $\tau_{13}$, $\tau_{23}$, and $\tau_{av}$ to be non-zero increases to $p_{\text{eff}} = 0.408$. From the effective noise parameter acting on qubit 3 while moving from the central node to the end node, we have found the number of repeaters and the total distance between the central and the end node. It is found that as the number of repeater nodes increases, the total as well the inter-repeater distances decrease. In case of third protocol, when we analyze how global entanglement of the final state evolves with noise, we find that unlike the other two protocols, the global entanglement between the qubits at the end nodes does not completely decay. Rather, at some value of the noise parameter $p$, it becomes minimum and remains constant thereafter. The reason behind this contrasting behaviour is that in the third protocol noise does not act directly on the qubits of the final shared W state. The resulting noise on those qubits is due to the three-qubit joint measurements on the qubits, which are subjected to noise. One line for a future study is to consider the distribution of more than three-qubit $W_{\mathrm{mod}}$ states. This will be a rich area with many versions of $W_{\mathrm{mod}}$ states and multiple protocols. We have also not considered the possibility of state distillation or quantum error corrections. Both processes will be integral parts of a future quantum network. If one could do state distillation at each repeater, then one can restore the $W_{\mathrm{mod}}$ state to its almost pure form at each repeater. This will increase the length of a network and possibility of implementing quantum communication protocols with higher figure of merits. Similarly, a suitable quantum error correction protocol can also help in restoring the state to its original form. These considerations can be a basis for future investigations. 

%\section*{Acknowledgements}

%\newpage

%\bibliographystyle{revtex4-2}
\bibliography{bibliography}

\newpage
\onecolumngrid
\appendix
\section*{Appendix}\label{appendix}
\section{Distribution of $W_{\mathrm{mod}}$ states using entanglement swapping}
In the main text, we have described the distribution of three-qubit $W_{\mathrm{mod}}$ states through teleportation and multipartite joint measurements. The underlying mechanism of these protocols is basically entanglement swapping. Here we analytically show how a locally prepared multipartite state can be distributed in a network using entanglement swapping.

\subsection{Bell-basis measurement}\label{appendix A1}
The second distribution protocol, as described in Section (\ref{sec3}), starts with the local preparation of a $W_{\mathrm{mod}}$ state. Let us consider that the prepared state in the central node be,   
\begin{equation}
\ket{W_{\mathrm{mod}}}_{123}=\frac{1}{2}\ket{100}_{123}+\frac{1}{2}\ket{010}_{12 3}+\frac{1}{\sqrt{2}}\ket{001}_{123},
\end{equation}
Consider the Bell state shared between the central node and one end node be,
\begin{equation}
\ket{\Phi^{+}}_{45}=\frac{1}{\sqrt{2}}(\ket{00}_{45}+\ket{11}_{45}).
\end{equation}
Therefore, the joint 5-qubit state can be expressed as,
\begin{eqnarray}\label{eq:psi0}
\ket{\Psi_0}_{12345}
&=& \ket{W_{\mathrm{mod}}}_{123}\otimes\ket{\Phi^+}_{45} \nonumber\\
&=& \frac{1}{2\sqrt{2}}\big( \ket{10000} + \ket{10011} + \ket{01000} + \ket{01011}\big) \nonumber\\
&&+ \frac{1}{2}\big( \ket{00100} + \ket{00111} \big).
\end{eqnarray}
Now, as the protocol describes, the central node transmits the qubits~1 and 2 to two end nodes. To distribute the target state across all the end nodes, a BSM is performed on the memory qubits~ 3 and 4, located at the central node. Grouping the terms  in the above equation using the Bell states $\ket{\Phi^\pm}=\frac{1}{\sqrt{2}}(\ket{00}\pm\ket{11})$ and $\ket{\Psi^\pm}=\frac{1}{\sqrt{2}}(\ket{01}\pm\ket{10})$ yields the following:
\begin{align}
\ket{\widetilde{\Psi}_{0}} 
&= \frac{1}{2}[\left( \frac{\ket{100}}{2} + \frac{\ket{010}}{2} + \frac{\ket{011}}{\sqrt{2}} \right)_{125}\ket{\Phi^+}_{34}
 + \left( \frac{\ket{100}}{2} + \frac{\ket{010}}{2} - \frac{\ket{001}}{\sqrt{2}} \right)_{125} \ket{\Phi^-}_{34} \nonumber \\
&\quad + \left( \frac{\ket{101}}{2} + \frac{\ket{011}}{2} + \frac{\ket{000}}{\sqrt{2}} \right)_{125} \ket{\Psi^+}_{34}
 - \left( \frac{\ket{101}}{2} + \frac{\ket{011}}{2} - \frac{\ket{000}}{\sqrt{2}} \right)_{125} \ket{\Psi^-}_{34}]
\end{align}
Using the identities,
\[
\ket{00}_{34}=\frac{\ket{\Phi^+}+\ket{\Phi^-}}{\sqrt{2}}~~ \text{and}\qquad
\ket{11}_{34}=\frac{\ket{\Phi^+}-\ket{\Phi^-}}{\sqrt{2}},
\]
and similarly for \(\ket{01} \text{and} \ket{10}\), we finally get the three-qubit state after the BSM outcome of $|\Phi^{+}\rangle$,

%Renormalizing (norm squared \(N = \langle\widetilde{\Psi}|\widetilde{\Psi}\rangle = 1\) for this idealized example) gives
\begin{equation}
\ket{\Psi_{125}}_{\Phi^+} = \frac{1}{2}\ket{100} + \frac{1}{2}\ket{010} + \frac{1}{\sqrt{2}}\ket{001}.
\end{equation}
This is the same functional form as \(\ket{W_{\mathrm{mod}}}\) but with qubit~5 replacing qubit~3; entanglement has been swapped from qubit~3 to qubit~5. For other Bell-state outcomes \(\{\Phi^-,\Psi^\pm\}\), the projected state on qubits (1,2,5) requires local Pauli corrections on qubit~5, which is located at one of the end nodes (or equivalently on one of the payload qubits). The appropriate corrections are determined from the specific Bell outcome and communicated classically to remote parties. In this way, using BSM-based entanglement swapping, a $W_{\mathrm{mod}}$ state can be distributed in an asymmetric star-shaped network.

%%%%%%%%%%%%%%%%%%%%%%%
\subsection{Three-qubit $W_{\mathrm{mod}}$-basis measurement}\label{appendix A2}

The distribution protocol for $W_{\mathrm{mod}}$ states involving multipartite joint measurements considers two distant central nodes $C$ and $C^{\prime}$ and an intermediary node $D$. Let us consider that the central nodes initially prepare the following states respectively, 
\begin{align}
\ket{W_{\mathrm{mod}}}_{123} &= \frac{1}{2}\ket{100}_{123}+\frac{1}{2}\ket{010}_{123}+\frac{1}{\sqrt{2}}\ket{001}_{123}, \\
\ket{W_{\mathrm{mod}}}_{456} &= \frac{1}{2}\ket{100}_{456}+\frac{1}{2}\ket{010}_{456}+\frac{1}{\sqrt{2}}\ket{001}_{456}.
\end{align}

The combined 6-qubit state is $\ket{\Psi_{\mathrm{tot}}}=\ket{W_{\mathrm{mod}}}_{123}\otimes\ket{W_{\mathrm{mod}}}_{456}$. In the second step of the protocol, the central node $C$ sends qubits~1 and 3 to two end nodes and $C^{\prime}$ sends qubit~6 to another end node. Once the central nodes complete the initial transmission of the respective qubits to the end nodes, node $C^{\prime}$ sends qubit~3, and $C^{\prime}$ sends qubits~5 and 6 to the intermediary node. The aim of the protocol is to distribute a $W_{\mathrm{mod}}$ state between three end nodes, where qubits~ 1, 2, and 6 have been transmitted. For this purpose, a joint three-qubit measurement on qubits (3, 4, 5) is performed at the intermediary node $D$ using an orthonormal $W_{\mathrm{mod}}$ basis set built with the following vectors:
\begin{align}
\ket{\eta_\pm} &= \frac{1}{2}\ket{010}+\frac{1}{2}\ket{001}\pm\frac{1}{\sqrt{2}}\ket{100}, \\
\ket{\zeta_\pm} &= \frac{1}{2}\ket{110}+\frac{1}{2}\ket{101}\pm\frac{1}{\sqrt{2}}\ket{000},
\end{align}
where the ordering of the qubits inside each ket is (3, 4, 5). These four vectors are mutually orthogonal, and together with four other orthogonal states (not used here), complete the 8-dimensional three-qubit space.

We expand \(\ket{\Psi_{\mathrm{tot}}}\) and collect terms according to the measurement on qubits (3, 4, 5). For clarity we write the  expanded product state (only terms relevant to \(\ket{\zeta_+}_{345}\) outcome are displayed here; the full expansion contains all 9 nonzero computational components before grouping) as:
\begin{align}
\ket{\Psi_{\mathrm{tot}}} &= \frac{1}{4}\ket{100\,100}
+ \frac{1}{4}\ket{100\,010}
+ \frac{1}{2\sqrt{2}}\ket{100\,001} \nonumber\\
&\quad + \frac{1}{4}\ket{010\,100}
+ \frac{1}{4}\ket{010\,010}
+ \frac{1}{2\sqrt{2}}\ket{010\,001} \nonumber\\
&\quad + \frac{1}{2\sqrt{2}}\ket{001\,100}
+ \frac{1}{2\sqrt{2}}\ket{001\,010}
+ \frac{1}{2}\ket{001\,001}.
\label{eq:tot6}
\end{align}
where the qubit ordering in each ket is \(1,2,3,4,5, \text{and}~6\) respectively.
Grouping qubits 3, 4, 5 and 1, 2, 6 separately, we get:
\begin{align}
 \ket{\Psi_{\mathrm{tot}}} &= \frac{1}{4}[\ket{010}_{345}\ket{100}_{126}+\ket{001}_{345}\ket{100}_{126}+\sqrt{2}\ket{000}_{345}\ket{100}_{126}
 \nonumber\\
 &\quad + \ket{010}_{345}\ket{010}_{126}+\ket{001}_{345}\ket{010}_{126}+\sqrt{2}\ket{000}_{345}\ket{011}_{126}
 \nonumber\\
 &\quad +\sqrt{2}\ket{000}_{345}\ket{011}_{126}+\sqrt{2}\ket{101}_{345}\ket{000}_{126}+2\ket{100}_{345}\ket{001}_{126}]   
\end{align}
Rewriting in terms of basis states yields:
\begin{align}
    \ket{\Psi_{\mathrm{tot}}}&=1/4[(\ket{\eta}_{+}+\ket{\eta}_{-})_{345}\ket{100}_{126}+(\ket{\eta}_{+}+\ket{\eta}_{-})_{345}\ket{010}_{126}+\sqrt{2}(\ket{\eta}_{+}-\ket{\eta}_{-})_{345}\ket{001}_{126}
    \nonumber\\
    &\quad (\ket{\zeta}_{+}-\ket{\zeta}_{-})_{345}(\ket{101}+\ket{011}_{126})+\sqrt{2}(\ket{\zeta}_{+}+\ket{\zeta}_{-})_{345}\ket{000}_{126}]
\end{align}

Finally, after re-arranging the terms, we get :
\begin{align}
    \ket{\Psi_{\mathrm{tot}}}&= 1/2[\ket{\eta+}_{345}(\ket{100}+\ket{010}+\sqrt{2}\ket{001})_{126}/2+\ket{\eta-}_{345}(\ket{000}+\ket{010}-\sqrt{2}\ket{001})_{126}/2 \nonumber\\
    &\quad +\ket{\zeta+}_{345}(\ket{101}+\ket{011}+\sqrt{2}\ket{000})_{126} 
  \ket{\zeta-}_{345}(-\ket{101}-\ket{011}+\sqrt{2}\ket{000})_{126})/2]
\end{align}
To project onto \(\ket{\eta_+}_{345}\), we compute,
\[
(\mathbb{I}_{12}\otimes\ket{\eta_+}\bra{\eta_+}_{345}\otimes\mathbb{I}_{6})\ket{\Psi_{\mathrm{tot}}}.
\]
After the projective measurement and tracing out of the measured qubits (3,4,5), the 3-qubit $W_{\mathrm{mod}}$ state that is shared between the end nodes is:
\[
\ket{\Phi_{126}} = \frac{1}{\sqrt{2}}\ket{001} + \frac{1}{2}\ket{010} + \frac{1}{2}\ket{100}.
\]
Different measurement outcomes from the set \(\{\eta_\pm,\zeta_\pm\}\) produce similar unitarily equivalent states, which can be corrected using local Pauli gates by communicating the measurement outcomes via classical channels.
%%%%%%%%%%%%%%%%%%%%%%%%%%%%%%%%%%%%%%%%%%%%%%%%%%%%%%%%%%

\subsection{Three-qubit general $\ket{W_m}$ state}\label{appendix A3}

We can do the analysis for the general state $\ket{W_m}$. This state, given in Eq. \ref{eq0}, can be deterministically generated by applying a two-qubit unitary between a Bell state and an ancilla qubit. From the product state
\begin{equation}\label{eq3}
|\phi^+\rangle_{13}\otimes|0\rangle_2 = \frac{1}{\sqrt{2}}\big(|000\rangle + |101\rangle\big),
\end{equation}
we can generate the state $\ket{W_m}$ by applying on qubits $(1,2)$ following unitary operation
\begin{equation}
U_{MWm} =
\begin{bmatrix}
0 & 0 & 1 & 0 \\
\tfrac{\sqrt{m}}{\sqrt{m+1}} & 0 & 0 & \tfrac{1}{\sqrt{m+1}} \\
\tfrac{1}{\sqrt{m+1}} & 0 & 0 & -\tfrac{\sqrt{m}}{\sqrt{m+1}} \\
0 & 1 & 0 & 0
\end{bmatrix}.
\end{equation}
%on qubits $(1,2)$ yields,
%\begin{equation}
%U_{MWn}(|\phi^+\rangle_{13}\otimes|0\rangle_2)
%= \frac{1}{\sqrt{2 + 2n}} (|100\rangle + \sqrt{n}|010\rangle + \sqrt{n + 1}|001\rangle),\nonumber\\
%\end{equation}
%which is exactly the desired modified W state. 

 We can also generate the measurement basis to implement the protocols for the state $\ket{W_m}$. For this we can apply the following unitary on qubits $(2,3)$ of the state in Eq (\ref{eq3}): 
 
\begin{equation}
U_{MWmB} =
\begin{bmatrix}
0 & 1 & 0 & 0 \\[4pt]
\tfrac{\sqrt{m}}{\sqrt{m+1}} & 0 & 0 & \tfrac{1}{\sqrt{m+1}} \\[4pt]
\tfrac{1}{\sqrt{m+1}} & 0 & 0 & -\tfrac{\sqrt{m}}{\sqrt{m+1}} \\[4pt]
0 & 0 & 1 & 0
\end{bmatrix}. \nonumber\\
\end{equation}
The resulting state will be,
\begin{eqnarray}
U_{MWmB}(|\phi^+\rangle_{13}\otimes|0\rangle_2) 
&=&  \tfrac{1}{\sqrt{2 + 2m}}(|010\rangle + \sqrt{m}|001\rangle + \sqrt{m + 1}|100\rangle) \nonumber\\
&=& |\eta^{+}_{m}\rangle,
\end{eqnarray}
which is one of the basis states. By applying local unitaries on qubit~1 of this state, the remaining three states $|\eta^{-}_{m}\rangle,|\xi^{+}_{m}\rangle,~ \text{and}~ |\xi^{-}_{m}\rangle$ of the set can be obtained as -- 
\[
|\eta^{-}_{m}\rangle = Z_1|\eta^{+}_{m}\rangle,\qquad
|\xi^{+}_{m}\rangle = X_1|\eta^{+}_{m}\rangle,\qquad
|\xi^{-}_{m}\rangle = X_1Z_1|\eta^{+}_{m}\rangle.
\]
where, $X$ and $Z$ are the Pauli operators.
Thus, the full orthonormal set $\{|\eta^{\pm}_{m}\rangle,|\xi^{\pm}_{m}\rangle\}$ can be realized from a single preparation circuit followed by simple, classically controlled single-qubit unitary operations. This construction provides a practical method for implementing $W_{\mathrm{mod}}$-basis measurements in network protocols. 

\section{Noise analysis}\label{appendix B}

In the main text, we utilize a simplified noise model where errors due to transmission, storage, and measurement are transferred to the final distributed qubits. While this technique is standard for stabilizer states (using the noisy stabilizer formalism), $W_{\mathrm{mod}}$ states do not belong to the stabilizer class. Therefore, we explicitly derive the validity of this approach here, relying on the linearity of quantum teleportation rather than Pauli group properties.\\
\newpage

\begin{proof}
\label{thm:commutativity}
\textit{Commutativity of Local Noise and Entanglement Swapping:}\\

Let the $W_{\mathrm{mod}}$ state be defined on qubits 1, 2, and 3 as:
\begin{equation}
    \ket{\psi}_{123} = \alpha\ket{100} + \beta\ket{010} + \gamma\ket{001}
\end{equation}
where qubit 3 is the memory qubit at the central node. Let $\mathcal{E}$ be a completely positive trace-preserving (CPTP) noise map acting on qubit 3. The process of subjecting qubit 3 to noise $\mathcal{E}$ followed by entanglement swapping to a remote node 5 is mathematically equivalent to performing ideal swapping followed by the application of $\mathcal{E}$ on the final node 5.
\end{proof}

\begin{proof}
Let the noise channel $\mathcal{E}$ be represented by Kraus operators $\{K_i\}$ such that $\mathcal{E}(\rho) = \sum_i K_i \rho K_i^\dagger$. 

The initial state of the system comprises the $W_{\mathrm{mod}}$ state (qubits 1,2,3) and a Bell pair $\ket{\Phi^+}_{45}$ shared between the central node (4) and the end node (5). The state after noise affects qubit 3 is:
\begin{equation}
    \rho_{total} = \sum_i \mathcal{K}_i \left( \ket{\psi}\bra{\psi}_{123} \otimes \ket{\Phi^+}\bra{\Phi^+}_{45} \right) \mathcal{K}_i^\dagger
\end{equation}
where $\mathcal{K}_i = I_{12} \otimes K_i^{(3)} \otimes I_{45}$.

We rewrite the $W_{\mathrm{mod}}$ state by isolating the basis of memory qubit 3. We define unnormalized vector components on qubits 1 and 2:
\begin{equation}
    \ket{v_0}_{12} = \alpha\ket{10} + \beta\ket{01}, \quad \ket{v_1}_{12} = \gamma\ket{00}
\end{equation}
Applying the $i$-th noise operator $K_i^{(3)}$ to the state yields:
\begin{equation}
    K_i^{(3)}\ket{\psi}_{123} = \ket{v_0}_{12} \otimes (K_i\ket{0})_3 + \ket{v_1}_{12} \otimes (K_i\ket{1})_3
\end{equation}

The entanglement swapping protocol constitutes a quantum teleportation channel $\mathcal{T}_{3 \to 5}$. By the linearity of quantum mechanics, this channel applies individually to the components of the superposition. For an ideal teleportation channel, $\mathcal{T}(\ket{\phi}_3) \mapsto \ket{\phi}_5$. Therefore:
\begin{align}
    \mathcal{T}_{3\to 5} \Big( \ket{v_0}_{12} \otimes (K_i\ket{0})_3 \Big) &= \ket{v_0}_{12} \otimes (K_i\ket{0})_5 \\
    \mathcal{T}_{3\to 5} \Big( \ket{v_1}_{12} \otimes (K_i\ket{1})_3 \Big) &= \ket{v_1}_{12} \otimes (K_i\ket{1})_5
\end{align}
Recombining these terms, the final trajectory for the $i$-th Kraus operator on qubits 1, 2, and 5 is:
\begin{align}
    \ket{\psi_{final}^{(i)}} &= I_{12} \otimes K_i^{(5)} \Big( \alpha\ket{100} + \beta\ket{010} + \gamma\ket{001} \Big)_{125} \nonumber \\
    &= (I_{12} \otimes K_i^{(5)}) \ket{\Tilde{W}}_{125}
\end{align}
Summing over all $i$, we recover the definition of the noise channel acting on qubit 5:
\begin{equation}
    \rho_{final} = (I_{12} \otimes \mathcal{E}_5) [\rho_{MW}]
\end{equation}
This proves that the noise is mapped isomorphically to the final qubit, preserving the coefficients $\alpha, \beta, \gamma$.
\end{proof}
\paragraph*{}
The theorem mentioned above assumes an ideal swapping channel. In physical implementations, the BSMs are noisy. We model a noisy BSM $\Lambda_{BSM}$ on qubits 3 and 4 as an ideal measurement preceded by local gate noise:
\begin{equation}
    \Lambda_{BSM} = \mathcal{M}_{ideal} \circ (\mathcal{E}_{gate}^{(3)} \otimes \mathcal{E}_{gate}^{(4)})
\end{equation}
Here, $\mathcal{E}_{gate}$ represents the depolarizing noise introduced by the CNOT and Hadamard gates within the BSM circuit.
From the above theorem, the gate noise $\mathcal{E}_{gate}^{(3)}$ acting on the memory qubit immediately before measurement commutes through the channel and appears on the target qubit 5. Since qubit 4 is part of the Bell pair $\ket{\Phi^+}_{45}$, we invoke the standard teleportation identity $(M \otimes I)\ket{\Phi^+} = (I \otimes M^T)\ket{\Phi^+}$. The gate noise $\mathcal{E}_{gate}^{(4)}$ is mathematically equivalent to a correlated noise channel acting on qubit 5. Thus, the effective noise on the final distributed state is the composition of all physical noise sources (memory, transmission, and BSM gates) accumulated onto the final node:
\begin{equation}
    \mathcal{E}_{eff}^{(5)} = \mathcal{E}_{BSM\_gate}^{(5)} \circ \mathcal{E}_{mem}^{(5)} \circ \mathcal{E}_{trans}^{(5)}
\end{equation}
This validates the simplified model used in the main text.\\

Using the above proof, when we analyze the performance of the second protocol in presence of noise, we find that two-tangle between qubits 1 and 2 decay faster than that between qubits 1(2) and 3. In FIG.~\ref{fig:12}, the evolution of the qubit-pairwise two-tangle are presented.

\begin{figure}
    \centering
    \begin{subfigure}{0.5\linewidth}
        \centering
        \includegraphics[width= \linewidth]{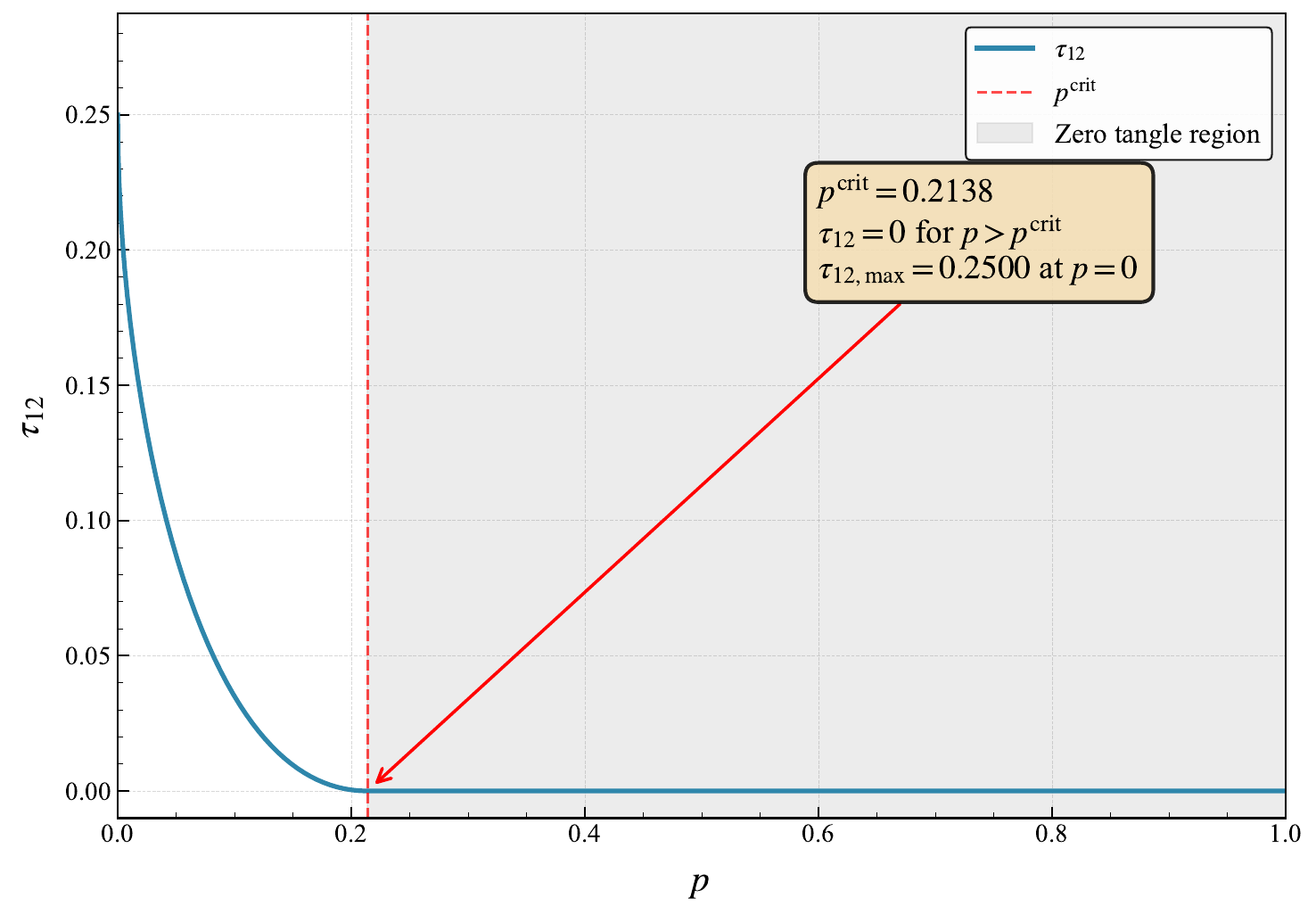}
        \caption{}
        \label{fig:12a}
    \end{subfigure}%
    \begin{subfigure}{0.5\linewidth}
        \centering
        \includegraphics[width= \linewidth]{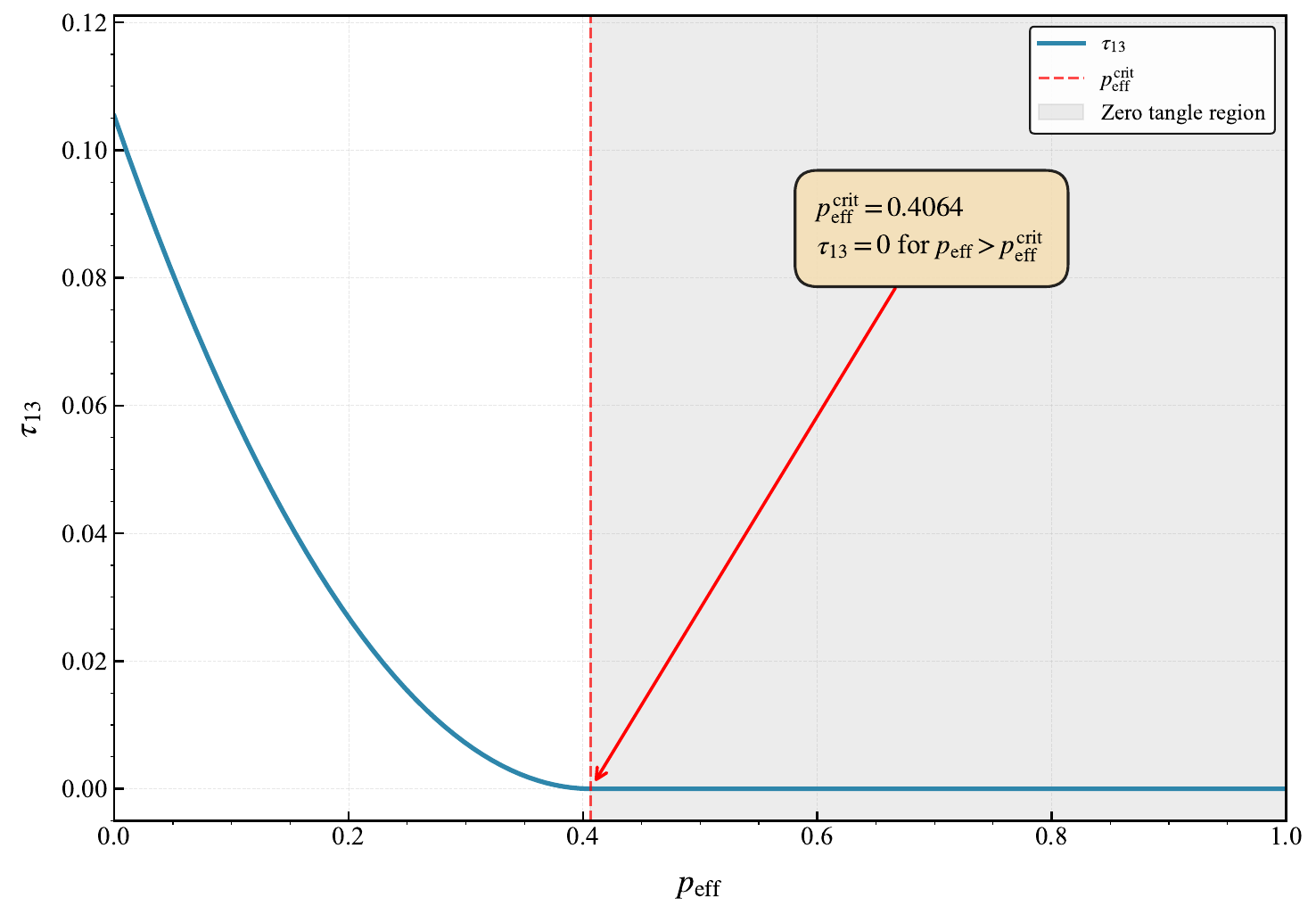}
        \caption{}
        \label{fig:12b}
    \end{subfigure}
    \caption {\justifying{Plots showing the variation of pairwise qubit entanglement with respect to the noise parameters in the second protocol. Figure (a) shows how two-tangle between qubits 1 and 2 changes with noise parameter $p$. Figure (b) shows the variation of two-tangle between qubits 1 (2)and 6 with noise.}}
    \label{fig:12}
\end{figure}

%\begin{figure}[h]
%    \centering
%   \includegraphics[width= 0.7\linewidth]{tau_av_measure_contour_final.pdf}
%   \caption{}
%   \label{fig:13}
%\end{figure}
%%%%%%%%%%%%%%%%%%%%%%%%%%%%%%%%%%%%%%%%%%%%%%%%%%%%%%%%%%

\end{document}